\documentclass[12pt,a4paper,onecolumn]{article}
\usepackage{jheppub}
\usepackage{hyperref}
\usepackage{amssymb}
\usepackage{amsmath}
\usepackage{appendix}
\usepackage{url}
\usepackage[most]{tcolorbox}
\usepackage{enumitem}
\usepackage{cancel}
\usepackage[absolute,overlay]{textpos}
\usepackage{xcolor} 
\newtcolorbox{eqhighlight}{colback=yellow!10!white, colframe=black, fonttitle=\bfseries, title=Notation}
\usepackage{textcomp}
\usepackage{tikz}
\usepackage{wrapfig}
\usepackage{stackengine}

\usepackage{amsmath}

\title{Metric-like Cubic Vertices for Massless Bosonic Higher-Spin Fields in AdS$_3$}

\author[a]{Freddie King,}
\author[b,c]{Taylor Pomfret}
\author[d]{and Karapet Mkrtchyan}

\affiliation[a]{Department of Physics and Astronomy, University of Sussex, BN1 9QH, Brighton, UK}
\affiliation[b]{Ludwig-Maximilians-Universit{\"a}t M{\"u}nchen, Geschwister-Scholl-Platz 1, D-80539 M{\"u}nchen, Germany}
\affiliation[c]{Max Planck Institute for Astrophysics, Karl-Schwarzschild-Stra{\ss}e 1, 85748 Garching bei M{\"u}nchen, Germany}
\affiliation[d]{Abdus Salam Centre for Theoretical Physics, Imperial College London, SW7 2AZ, United Kingdom}

\emailAdd{fk260@sussex.ac.uk}
\emailAdd{pomfret@mpa-garching.mpg.de}
\emailAdd{k.mkrtchyan@imperial.ac.uk}

\abstract{We derive parity-even transverse-traceless metric-like cubic vertices for massless bosonic higher-spin fields in AdS$_3$. Starting from the known two- and three-derivative vertices in three-dimensional flat space, we construct their AdS$_3$ extensions by imposing gauge invariance and accounting for dimension-dependent identities. The result is shown to be consistent with minimal coupling to gravity.
}

\begin{document}

{\phantom{.}\vspace{-2.5cm}\\\flushright Imperial-TP-KM-2026-02\\ }

\maketitle

\newpage

\section{Introduction}
\label{Introduction}
\indent Theories of higher-spin (HS) gravity, in which fields of spin greater than two couple to the gravitational field, are promising candidates for formulating a consistent description of quantum gravity. They naturally arise in the tensionless limit of string theory, which generates an infinite spectrum of massless integer HS fields \cite{Sundborg:2000wp,Bonelli:2003kh,Sagnotti:2003qa}. Additionally, models of HS gravity \cite{Vasiliev:1990en,Vasiliev:1995dn,Vasiliev:2003ev,Bekaert:2004qos,Didenko:2014dwa} are linked to well-understood conformal field theories via AdS/CFT \cite{Maldacena:1997re,Gubser:1998bc,Witten:1998qj}. An early example is the higher-spin/vector-model duality, put forward by Klebanov and Polyakov \cite{Klebanov:2002ja}, which relates higher-spin gauge theory in AdS$_4$ to critical large-$N$ $O(N)$ vector models on the boundary; see \cite{Konstein:2000bi,Sezgin:2002rt,Manvelyan:2004ii,Manvelyan:2008ks,Giombi:2011ya,Maldacena:2011jn,Giombi:2012ms,Giombi:2013yva,Giombi:2013fka,Giombi:2014iua,Giombi:2014yra,Costa:2014kfa,deMelloKoch:2014vnt,Bekaert:2014cea,Beccaria:2015vaa,Beccaria:2016tqy,Sleight:2016dba,Bae:2016rgm,Bae:2016hfy,Gunaydin:2016amv,Giombi:2016pvg,Giombi:2016zwa,Bae:2017spv,Didenko:2017lsn,Bae:2017fcs,Basile:2018zoy,Fredenhagen:2018guf,Aharony:2020omh,Neiman:2022enh} for a non-exhaustive list of works on related developments and \cite{Giombi:2016ejx,Bekaert:2022poo} for reviews. The three-dimensional case, $AdS_3/CFT_2$ 
\cite{Gaberdiel:2010pz,Gaberdiel:2012uj}, holds particular interest because the boundary two-dimensional CFT possesses an infinite-dimensional conformal symmetry \cite{Brown:1986nw,Campoleoni:2010zq,Henneaux:2010xg,Campoleoni:2011hg}, related to known theories \cite{Belavin:1984vu}, allowing for more general, exact results.\\
\indent The metric-like approach, based on Fronsdal's original formulation \cite{Fronsdal:1978rb,Fang:1978wz}, is built from symmetric tensor fields, in analogy to the metric-based description familiar from standard general relativity. In the free theory of HS massless bosonic fields, the Fronsdal equation can be seen as a generalisation of the Maxwell and (linearised) Einstein field equations to arbitrary integer spin-$s$, see e.g. \cite{Kessel:2017mxa,Bekaert:2022poo,Ponomarev:2022vjb,Pekar:2023nev} for reviews. A key challenge is the absence of a complete metric-like action for the fully interacting theory \cite{Bekaert:2010hw,Bekaert:2004qos}. The Fronsdal program addresses this by perturbative construction of the Lagrangian, starting from the free Fronsdal action \cite{Fronsdal:1978rb} and constructing interactions order by order using the Noether procedure \cite{Berends:1984rq}. The first non-trivial step in this process is the classification of cubic vertices.\footnote{In fact, higher-order independent vertices (i.e. vertices that come with new coupling constants independent of the cubic ones) are absent in $d=3$, as shown in \cite{Fredenhagen:2019hvb}.}\\
\indent Three-dimensional flat spacetime is particularly attractive \cite{Campoleoni:2024ced} because the typical obstructions to minimal gravitational coupling, namely the Weinberg S-matrix argument \cite{Weinberg:1964ew} and the Aragone–Deser problem, do not apply \cite{Aragone:1979hx, Aragone:1983sz}. In particular, dimension-dependent identities (DDIs) in $d=3$ make it possible to relax the usual gauge symmetry constraints, leading to two- and three-derivative cubic vertices for massless bosonic fields \cite{Mkrtchyan:2017ixk,Kessel:2018ugi}, including minimal coupling to gravity. The same identities are responsible for trivialising candidate vertices involving higher derivatives. This is in contrast to general dimensions, where the number of derivatives grows with the highest spin in the vertex\footnote{Note that the number of derivatives is a good identifier of vertices in the metric-like description in flat space, since a given triplet of spins and number of derivatives has at most one vertex. The frame-like classification of vertices goes in parallel, though the number of derivatives is not transparent there without explicit solution to the torsion constraints \cite{Vasiliev:2011knf,Boulanger:2012dx}. Furthermore, the $(A)dS$ vertices contain terms with different number of derivatives, but they can be viewed as deformations of flat space vertices and are also given by a unique linearly independent basis parameterised by the spins and the highest number of derivatives involved.} \cite{Bengtsson:1983pd,Metsaev:2005ar,Metsaev:2007rn,Manvelyan:2010jr}. In particular, in $d\ge4$, the two-derivative minimal gravitational coupling is not available for Fronsdal fields \cite{Boulanger:2008tg} of spin greater than two\footnote{Note that four dimensions are special: here, minimal coupling to gravity exists in the light-cone formalism \cite{Bengtsson:1986kh,Metsaev:1991mt,Metsaev:2018xip} and at the amplitude level \cite{Benincasa:2007xk,Benincasa:2011pg,Conde:2016vxs,Conde_2016} (see also \cite{Khabarov:2020bgr}).}. Another related special feature of three dimensions is the Chern-Simons formulation, originally proposed for $AdS_3$ gravity \cite{Achucarro:1986uwr,Witten:1988hc} and soon after extended to higher spins \cite{Blencowe:1988gj,Vasiliev:1989re,Campoleoni:2010zq}. While the symmetry algebras are better understood in constant curvature spacetimes, field theory is more tractable in flat spacetime. Fortunately, the flat space limits of three-dimensional higher spin theories are regular \cite{Afshar:2013vka,Grigoriev:2020lzu,Boulanger:2023tvt}, and the flat space interactions can be studied as the first step towards the full $AdS_3$ theory.\footnote{Despite original indications of the special role of $AdS$ in higher-spin gravity \cite{Vasiliev:1986bq,Fradkin:1987ks,Fradkin:1986qy}, many features of interactions are very similar between flat and $AdS$ spaces \cite{Boulanger:2008tg,Manvelyan:2010jr,Joung:2011ww}, even in $d\geq 4$. The main difference is in gauge and global symmetry structures. See \cite{Joung:2013nma} for the metric-like analogue of Fradkin-Vasiliev mechanism of regaining diffeomorphism symmetry (absent in flat space due to Aragone-Deser problem, sourcing no-go theorems \cite{Porrati:2012rd}) via non-zero cosmological constant.}

In the Chern-Simons formulation, the massless fields of spin greater than or equal to two carry no local propagating degrees of freedom. They are given by a short representation of the $AdS_3$ isometry algebra $\mathfrak{so}(2,2)
\sim \mathfrak{so}(1,2)\oplus \mathfrak{so}(1,2)$, where one of the simple components carries a trivial representation (see, e.g., \cite{Gwak:2015jdo,Alkalaev:2021zda}).
The cubic vertices of massless higher spins in three dimensions should therefore not be interpreted as ordinary bulk scattering data. Instead, they provide local representatives of the interaction and gauge-structure data underlying the global, boundary, and asymptotic symmetry structures of the theory \cite{Brown:1986nw,Campoleoni:2010zq,Campoleoni:2012hp,Fredenhagen:2018guf}. The broader landscape of these theories is further fleshed out by fully non-linear higher-spin constructions in AdS$_3$, such as Blencowe's Chern-Simons action for interacting massless fields of all integer and half-integer spins \cite{Blencowe:1988gj}, and the Vasiliev and Prokushkin-Vasiliev equations, including matter-coupled systems \cite{Vasiliev:1992gr,Vasiliev:1990en,Vasiliev:1992ix,Prokushkin:1998vn}. These results are strong evidence that the known flat space vertices \cite{Mkrtchyan:2017ixk} should admit gauge-invariant AdS$_3$ extensions - an expectation we confirm by explicitly deriving the two- and three-derivative AdS$_3$ cubic vertices, which reduce to the flat space counterparts as the radius of curvature becomes large.\\
\indent Although all parity-even cubic interactions of massless bosonic fields in flat three-dimensional spacetime have been constructed \cite{Mkrtchyan:2017ixk}, their AdS counterparts are not known in the metric-like formulation. In this work, we start from the known flat space vertices and add the most general lower-derivative AdS corrections compatible with the operator structure, which are then fixed by the requirement of gauge invariance. The introduction of a background with non-zero curvature introduces many subtleties in the computations due to non-commuting derivatives, requiring a more general procedure than in the flat case.\\
\indent This paper is organised as follows. Section~\ref{Preliminaries: free higher-spin theory and metric-like cubic vertices in three-dimensional flat space} reviews free higher-spin theory and the metric-like formulation of cubic interactions in three-dimensional flat space. In Section~\ref{Metric-like cubic vertices in AdS_3}, we extend this framework to AdS$_3$. This includes the construction of an appropriate basis of DDIs in AdS and the derivation of the curvature corrections to the known flat space vertices. We also perform an important consistency check by recovering the minimal gravitational coupling obtained from the Fronsdal action. Section~\ref{Summary and Conclusions} summarises the main results and discusses their implications. Technical details are collected in the appendices.
\section{Preliminaries: free theory and flat space vertices in $d=3$}
\label{Preliminaries: free higher-spin theory and metric-like cubic vertices in three-dimensional flat space}
\subsection{The Fronsdal Equation}
\label{The Fronsdal Equation}
In Fronsdal’s formulation \cite{Fronsdal:1978rb}, one represents a bosonic spin-s field as a fully symmetric rank-$s$ tensor $\phi_{\mu_1...\mu_s}$. Fronsdal's equation, which can be viewed as a generalisation of the Maxwell and linearised Einstein equations \cite{Kessel:2017mxa}, is given by\footnote{In our convention, parentheses around indices denote symmetrisation with weight one.}\footnote{One usually supplements equation \eqref{fronsdal equation} with the double-tracelessness condition
$\phi^{\rho\sigma}{}_{\rho\sigma\mu_1\cdots\mu_{s-4}}=0$. This constraint is sufficient, but not necessary, for the propagation of the correct degrees of freedom, and is useful as it allows us to obtain a gauge-invariant Fronsdal action; see, e.g., \cite{Kessel:2017mxa,Ponomarev:2022vjb}.}
\begin{equation}
\begin{aligned}
F_{\mu_1...\mu_s} = \Box \phi_{\mu_1...\mu_s} - s\partial_{(\mu_1}\partial^{\sigma}\phi_{\mu_2...\mu_s)\sigma} + \frac{s(s-1)}{2}\partial_{(\mu_1}\partial_{\mu_2}\phi_{\mu_3...\mu_s)\sigma}^{\hspace{1.2cm}\sigma} = 0,
\end{aligned}
\label{fronsdal equation}
\end{equation}
which has a gauge symmetry
\begin{equation}
\delta\phi_{\mu_1...\mu_s} = \partial_{(\mu_s}\xi_{\mu_1...\mu_{s-1})},
\label{fronsdal gauge symmetry}
\end{equation}
provided the gauge parameter, $\xi_{\mu_1...\mu_{s-1}}$ is traceless in any pair of indices.
A natural attempt to extend the Fronsdal equation \eqref{fronsdal equation} to a non-flat background would be to covariantise the derivatives, but because covariant derivatives do not commute, this would not preserve the gauge symmetry. For constant-curvature backgrounds, this is fixed by the inclusion of the two additional highlighted terms \cite{Kessel:2017mxa,Fronsdal:1978vb}
\begin{multline}
F^{\mu_1...\mu_s} = \Box\phi^{\mu_1...\mu_s} - s\nabla^{(\mu_1}\nabla_{\nu}\phi^{\mu_2...\mu_s)\nu} + \frac{s(s-1)}{2}\nabla^{(\mu_1}\nabla^{\mu_2}{\phi^{\mu_3...\mu_s)\nu}}_{\nu}\\
\colorbox{gray!15}{$- \frac{1}{\ell^2}[(s-2)(d+s-3)-s]\phi^{\mu_1...\mu_s} - \frac{1}{\ell^2}s(s-1)g^{(\mu_1\mu_2}{\phi^{\mu_3...\mu_s)\nu}}_{\nu}$}.
\label{AdS Fronsdal equation}
\end{multline}
By fixing the gauge, one can arrive at a Fierz-Pauli-like transverse-traceless (TT) system \cite{Fierz:1939ix,Ponomarev:2022vjb} (also known as Fierz system \cite{Fierz:1939zz}):
\begin{align}
\left(\Box - \frac{1}{\ell^2}[(s-2)(s+d-3)-s]\right)\phi^{\mu_1...\mu_s} &= 
0,
\label{AdS TT gauge on-shell conditions 1}\\
\nabla_{\nu}\phi^{\nu\mu_1...\mu_{s-1}} &= 0,
\label{AdS TT gauge on-shell conditions 2}\\
{\phi_{\nu}}^{\nu\mu_1...\mu_{s-2}} &= 0,
\label{AdS TT gauge on-shell conditions 3}
\end{align}
which has residual gauge symmetry
\begin{equation}
\delta\phi^{\mu_1...\mu_s} = \nabla^{(\mu_s}\xi^{\mu_1...\mu_{s-1})},
\label{AdS gauge symmetry}
\end{equation}
provided the gauge parameter $\xi^{\mu_1...\mu_{s-1}}$ satisfies
\begin{align}
\left(\Box - \frac{1}{\ell^2}(s-1)(s+d-3)\right)\xi^{\mu_1...\mu_{s-1}} &= 
0,
\label{AdS TT gauge parameter conditions 1}\\
\nabla_{\nu}\xi^{\nu\mu_1...\mu_{s-2}} &= 0,
\label{AdS TT gauge parameter conditions 2}\\
{\xi_{\nu}}^{\nu\mu_1...\mu_{s-3}} &= 0.
\label{AdS TT gauge parameter conditions 3}
\end{align}
Although in practice we do not gauge fix when analysing higher-order terms in the action, the structure of these equations is important for handling d’Alembertians in the on-shell TT gauge variation to be employed in the Noether procedure in Section~\ref{The Noether Procedure}.
\subsection{Cubic Vertices From a Generating Function}
\label{Cubic Vertices From a Generating Function}
We now briefly review the classification of parity-even massless bosonic cubic vertices in three dimensions, as originally set out in \cite{Mkrtchyan:2017ixk}. A similar method has been used for the classification of parity-odd and Chern-Simons vertices \cite{Kessel:2018ugi}, as well as vertices involving massless fermionic Fang-Fronsdal fields  \cite{Fredenhagen:2024lps}, but these cases lie outside the present scope.\\
\indent The cubic action can be written schematically as
\begin{equation}
\mathcal{S}_3 = \int d^d x\,\sum_{s_1\geq s_2\geq s_3}\sum_n g^{\,n}_{s_1,s_2,s_3}\,
\mathcal{L}^{(n)}_{s_1,s_2,s_3},
\label{eq:general-cubic-lagrangian}
\end{equation}
where \(n\) labels the independent cubic structures for a given spin triple ($s_1\geq s_2\geq s_3$ without loss of generality), and \(g^{\,n}_{s_1,s_2,s_3}\) are the corresponding coupling constants. Vertex construction is significantly simplified by using index-free generating functions for the fields and gauge parameters
\cite{Manvelyan:2010je,Manvelyan:2010jr,Joung:2011ww,Joung:2012fv,Mkrtchyan:2017ixk,Fredenhagen:2024lps,Kessel:2018ugi},
\begin{align}
\phi_i(x_i,a_i) &= \frac{1}{s_i!}\,
\phi_{\mu_1\cdots\mu_{s_i}}(x_i)
a_i^{\mu_1}\cdots a_i^{\mu_{s_i}},
\label{eq:field-generating-function}\\
\xi_i(x_i,a_i) &= \frac{1}{(s_i-1)!}\,
\xi_{\mu_1\cdots\mu_{s_i-1}}(x_i)
a_i^{\mu_1}\cdots a_i^{\mu_{s_i-1}}.
\label{eq:gauge-param-generating-function}
\end{align}
This notation avoids cumbersome explicit index contractions while making the complete symmetrisation of the fields manifest. In terms of the fields, one can write \eqref{eq:general-cubic-lagrangian} as
\begin{equation}
S_3 = 
\left.
  \int d^dx \prod_{j=1}^3 dx_j \,\delta(x - x_j)
  \sum_{s_1\geq s_2\geq s_3}\sum_n\mathcal{V}^{(n)}_{s_1,s_2,s_3}\,\phi_1\,\phi_2\,\phi_3
\right|_{a_i = 0},
\label{making cubic action local}
\end{equation}
where $\mathcal{V}^{(n)}_{s_1,s_2,s_3}$ is a vertex operator, in principle depending on derivatives with respect to $x_i$ or $a_i$. A set of Lorentz invariant operators that span the space of all possible field combinations is given by
\begin{align} 
&P_{ij} = \nabla_i\cdot\nabla_{j}, \hspace{0.5cm} y_i = \partial_i\cdot\nabla_{i+1}, \hspace{0.5cm} y_i' = \partial_i\cdot\nabla_{i-1}, \hspace{0.5cm} \nonumber \\& z_i = \partial_{i+1}\cdot\partial_{i-1}, \hspace{0.5cm} \mathrm{Tr}_i = \partial_i\cdot\partial_i, \hspace{0.5cm} \mathrm{Div}_i = \partial_i\cdot\nabla_i,
\label{naive operator basis}
\end{align}
where ${\nabla_i}_{\mu} \equiv \frac{\partial}{\partial x_i^{\mu}}, \hspace{0.5cm} \partial_{i\mu} \equiv \frac{\partial}{\partial a_i^{\mu}}$, and $i\in\{1,2,3\}$ is cyclic modulo three. Assuming one can discard total derivatives in the action, integration by parts (IBP) reveals that \eqref{naive operator basis} is over-complete. In particular, one finds
\begin{align}
y_i' &\overset{\text{IBP}}{\sim}y_i + \mathrm{Div}_i,
\label{y_i integration by parts}\\
P_{ii+1} &\overset{\text{IBP}}{\sim}\frac{1}{2}\left(\Box_{i-1} - \Box_i - \Box_{i+1}\right).
\label{Pii+1 integration by parts}
\end{align}
The second relation \eqref{Pii+1 integration by parts} reduces all two-derivative structures to d'Alembertians, which can then be removed up to traces and divergences by field redefinitions.\footnote{Schematically, consider a field redefinition of the form
\begin{equation}
\phi_1 \mapsto \phi_1 + g\phi_2\phi_3
\label{field redefinition example}
\end{equation}
which transforms the Fronsdal action (written now for three field copies),
\begin{align}
 S_2 \sim \frac{1}{2}\int d^dx&\left(\sum_i\phi_i\Box\phi_i + \text{tr/div terms} \right)\nonumber\\ &\mapsto S_2 + \frac{g}{2}\int d^dx\left(\phi_2\phi_3\Box\phi_1 + \text{cubic tr/div terms} + ...\right),
\label{field redef new vertices}
\end{align}
such that terms containing $\phi_1$ can generate new cubic contributions at $\mathcal{O}(g)$. This effectively transfers d’Alembertians, traces, and divergences into the cubic action, meaning that it is possible to trade d’Alembertian terms with trace/divergence terms through suitable field redefinitions.} 
Lastly, traces and divergences may be discarded if one is interested only in the classification of independent vertices rather than a full off-shell completion. The full cubic vertices can then be reconstructed order by order in traces and divergences, starting from the TT part and imposing gauge invariance \cite{Manvelyan:2010jr,Manvelyan:2010je,Francia:2016weg}. In this sense, the classification of TT vertices amounts to solving the next-order deformation of the Fierz system \eqref{AdS TT gauge on-shell conditions 1}--\eqref{AdS TT gauge on-shell conditions 3} \cite{Fierz:1939ix,Fierz:1939zz}.\\
\indent In summary, independent TT cubic vertices are classified by operators built from the basic contractions \(y_i\) and \(z_i\). The most general TT vertex can therefore be encoded as $\mathcal{V}(y_i,z_i)$.
\subsection{The Noether Procedure}
\label{The Noether Procedure}
\indent Implicit in our construction is that there exists some fully interacting action S that can be written as an expansion in powers of some coupling constant $g$:
\begin{equation}
S = S_2 + gS_3 + g^2S_4 + ...,
\label{non linear action expansion}
\end{equation}
with $S_2$ referring to the free Fronsdal action, while $S_3$ is cubic in fields, $S_4$ is quartic and so on. One can similarly expand deformations of the gauge symmetry,
\begin{equation}
\delta^{\xi}\phi = \delta_0^{\xi}\phi + g\delta_1^{\xi}\phi + g^2\delta_2^{\xi}\phi + ...,
\label{gauge symmetry expansion}
\end{equation}
where $\delta_0^{\xi}$ is the free Fronsdal gauge symmetry, corresponding to (\ref{fronsdal gauge symmetry}) or (\ref{AdS gauge symmetry}), while $\delta_1^{\xi}$ and $\delta_2^{\xi}$ are higher order corrections that are linear and quadratic in $\phi$ respectively.\footnote{Note that we can keep the gauge transformations linear in $\xi$ via redefinition of the gauge parameter, see e.g. \cite{Ponomarev:2022vjb}}
Gauge invariance of the full action requires
\begin{equation}
\delta S = 0,
\label{eq:full-nonlinear-gauge-symmetry}
\end{equation}
which implies the following on-shell constraint on the cubic vertices at \(\mathcal{O}(g)\):
\begin{equation}
\delta_0^{\xi}S_3 \approx 0.
\label{eq:cubic-vertex-gauge-symmetry}
\end{equation}
Here $\approx$ denotes equality modulo the free equations of motion. To translate this condition into the generating-function language of
\eqref{eq:field-generating-function} and \eqref{eq:gauge-param-generating-function},
one first realises that the gauge variation of a field \eqref{fronsdal gauge symmetry},\eqref{AdS gauge symmetry} can be written
\begin{equation}
\delta_i \phi_i = a_i\cdot\nabla_i \xi_i,
\label{eq:field-gauge-variation-gen-func}
\end{equation}
where $\delta_i$ denotes the lowest-order gauge variation $\delta_0^\xi$ acting on the $i$th field in the cubic vertex. Then the gauge variation of a general cubic vertex reads
\begin{equation}
\delta_i\mathcal{L} = \mathcal{V}(y_i,z_i)\phi_{i-1}\left(a_i\cdot\nabla_i\xi_i\right)\phi_{i+1},
\label{gauge variation of cubic vertex}
\end{equation}
which can then be recast in terms of a commutator,
\begin{equation}
[\mathcal{V}, \left(a_i\cdot\nabla_i\right)]\phi_{i-1}\xi_i\phi_{i+1},
\label{gauge variation commutator}
\end{equation}
because the term with $\left(a_i\cdot\nabla_i\right)$ on the left of the vertex operator $\mathcal{V}$ contains a residual auxiliary vector, meaning it is automatically set to zero in \eqref{making cubic action local}. Simplifying \eqref{gauge variation commutator} requires the commutators of $\left(a_i\cdot\nabla_i\right)$ with the $y_i$ and $z_i$. These are straightforward to compute in flat spacetime:
\begin{align}
[y_i, \left(a_i\cdot\nabla_i\right)] = P_{ii+1} \approx 0, \hspace{0.5cm} [y_{i\pm1}, \left(a_i\cdot\nabla_i\right)] = 0,\nonumber\\
[z_i, \left(a_i\cdot\nabla_i\right)] = 0, \hspace{0.5cm}[z_{i\pm1},\left(a_i\cdot\nabla_i\right)] = \pm y_{i\mp1}.
\label{y and z commutators}
\end{align}
Under the natural assumption that $\mathcal{V}$ is polynomial in $(y_i,z_i)$, and using the Leibniz property of the commutator, the relations \eqref{y and z commutators} imply
\begin{equation}
\delta_i\mathcal{V} = \left(y_{i-1}\partial_{z_{i+1}} - y_{i+1}\partial_{z_{i-1}}\right)\mathcal{V} = 0, \hspace{0.5cm} \forall i\in\{1,2,3\},
\label{differential gauge variation on general vertex}
\end{equation}
providing a useful differential formulation of the gauge symmetry condition \eqref{eq:cubic-vertex-gauge-symmetry}. Although this condition is readily solved in arbitrary dimension, the resulting basis does not contain the ordinary two-derivative minimal gravitational coupling for spins $s>2$, and four-dimensional DDIs merely reduce this basis without producing new vertices \cite{Manvelyan:2010je,Manvelyan:2010jr,Mkrtchyan:2017ixk}.
\subsection{Dimension-dependent Identities}
\label{Dimension-dependent Identities}
The crucial aspect of working in three dimensions is that there are non-trivial combinations of the vertex operators that vanish identically, known as DDIs or Schouten identities, that expand the space of solutions to \eqref{differential gauge variation on general vertex}. These can generally be obtained by antisymmetrising over $d+1$ indices in $d$ dimensions. The key observation is that the condition of gauge invariance \eqref{differential gauge variation on general vertex} can be recast as
\begin{equation}
\delta_i\mathcal{V} = \left(y_{i-1}\partial_{z_{i+1}} - y_{i+1}\partial_{z_{i-1}}\right)\mathcal{V} \propto \mathrm{DDIs}, \hspace{0.5cm} \forall i\in\{1,2,3\},
\label{gauge invariance up to DDIs}
\end{equation}
because the DDIs vanish identically. To systematically derive the Schouten identities in three dimensions in terms of the $y_i$ and $z_i$, one contracts sets of eight basic operators $\nabla_i$ and $\partial_i$, then antisymmetrises over four distinct indices. Again, we discard d'Alembertians, traces and divergences, since such terms only interact with the gauge variation of the non-TT components of the completed vertex. One can readily show that the Schouten identities can only contain two, three or four derivatives, i.e. two, three, or four powers of the $y_i$.\footnote{A non-trivial contracted set must contain at least one of any basic operator type $\nabla$, $\partial$ in each of the two groups of four operators being antisymmetrised over, otherwise it is guaranteed that at least one operator will be repeated among the antisymmetrised indices, causing it to vanish. Additionally, a non-trivial set cannot contain more than four derivatives; any additional derivatives would render the expression trivially zero or equivalent to a total derivative.} In total, there are eighteen independent identities \cite{Mkrtchyan:2017ixk}, which can be written in a reduced form due to cyclic symmetry:
\begin{align}
\nabla_{i[\mu}\partial_{i\nu}\partial_{i+1\rho}\partial_{i-1\sigma]}
\nabla_i^\mu\partial_i^\nu\partial_{i+1}^\rho\partial_{i-1}^\sigma
&\propto (G - y_iz_i)^2 \equiv 0,
\label{eq:(G - y_iz_i)^2 flat-space-schouten-identities}\\
\nabla_{i+1[\mu}\partial_{i\nu}\partial_{i+1\rho}\partial_{i-1\sigma]}
\nabla_{i-1}^\mu\partial_i^\nu\partial_{i+1}^\rho\partial_{i-1}^\sigma
&\propto y_iz_iG - y_{i+1}z_{i+1}y_{i-1}z_{i-1} \equiv 0,
\label{eq:y_iz_iG - yi+1zi+1yi-1zi-1flat-space-schouten-identities}\\
\nabla_{i[\mu}\nabla_{i+1\nu}\partial_{i\rho}\partial_{i-1\sigma]}
\nabla_i^\mu\partial_i^\nu\partial_{i+1}^\rho\partial_{i-1}^\sigma
&\propto y_iy_{i-1}(G - y_iz_i) \equiv 0,
\label{eq:yiyi-1(G - y_iz_i)flat-space-schouten-identities}\\
\nabla_{i-1[\mu}\nabla_{i\nu}\partial_{i\rho}\partial_{i+1\sigma]}
\nabla_i^\mu\partial_{i-1}^\nu\partial_i^\rho\partial_{i+1}^\sigma
&\propto y_iy_{i+1}(G - y_iz_i) \equiv 0,
\label{eq:y_iy_i+1(G - y_iz_i) flat-space-schouten-identities}\\
\nabla_{i[\mu}\nabla_{i+1\nu}\partial_{i+1\rho}\partial_{i\sigma]}
\nabla_{i+1}^\mu\nabla_{i-1}^\nu\partial_i^\rho\partial_{i+1}^\sigma
&\propto y_i^2y_{i+1}^2 \equiv 0,
\label{eq:y_i^2y_i+1^2 flat-space-schouten-identities}\\
\nabla_{i-1[\mu}\nabla_{i\nu}\partial_{i\rho}\partial_{i-1\sigma]}
\nabla_{i-1}^\mu\nabla_i^\nu\partial_i^\rho\partial_{i+1}^\sigma
&\propto y_i^2y_{i+1}y_{i-1} \equiv 0.
\label{eq:y_i^2y_i+1y_i-1 flat-space-schouten-identities}
\end{align}
where we have defined
\begin{align}
G \equiv \sum_i y_iz_i.
\label{G definition}
\end{align}
Together with cyclic permutations, these identities form a complete basis for the operator relations generated by three-dimensional Schouten identities.
\subsection{Classification by number of derivatives}
\label{Classification by number of derivatives}
Because the three-dimensional Schouten identities strongly constrain the derivative order, the classification is naturally organised by the number of derivatives. The genuinely new higher-spin structures occur at two and three derivatives. For two-derivative vertices, complete index contraction requires $s_1+s_2+s_3$ to be even. The non-trivial family exists when the strict triangle inequalities are satisfied,
\begin{align}
s_{i-1}+s_{i+1}\geq s_i+2,
\label{strict triangle inequality 2 deriv vertex}
\end{align}
where the shift by two reflects the two derivatives carried by the vertex. It is given by
\begin{align}
\mathcal{V}^{(2)}_{s_1,s_2,s_3}
=\left[(s_1-1)y_1z_1+(s_2-1)y_2z_2+(s_3-1)y_3z_3\right]Gz_1^{n_1}z_2^{n_2}z_3^{n_3},
\label{eq:flat-two-derivative-vertex}\\
n_i=\frac12(s_{i-1}+s_{i+1}-s_i)-1\geq0,
\label{eq:flat-two-derivative-ni}
\end{align}
which is gauge invariant by virtue of \eqref{eq:y_iz_iG - yi+1zi+1yi-1zi-1flat-space-schouten-identities}, \eqref{eq:yiyi-1(G - y_iz_i)flat-space-schouten-identities} and \eqref{eq:y_iy_i+1(G - y_iz_i) flat-space-schouten-identities}.
This family contains the two-derivative minimal gravitational coupling: setting $s_1=s_2=s$ and $s_3=2$ gives the $s$-$s$-$2$ vertex expected from a metric perturbation in the free Fronsdal action up to equivalence under Schouten identities. For three-derivative vertices, complete contraction instead requires \(s_1+s_2+s_3\) to be odd. Again, the non-trivial family exists when triangle inequalities are satisfied, now requiring
\begin{align}
s_{i-1}+s_{i+1}>s_i.
\label{eq:strict triangle inequality 3 deriv vertex}
\end{align}
We then find
\begin{align}
\mathcal{V}^{(3)}_{s_1,s_2,s_3} = y_1y_2y_3z_1^{p_1}z_2^{p_2}z_3^{p_3},
\label{eq:flat-three-derivative-vertex}\\
p_i = \frac12(s_{i-1}+s_{i+1}-s_i-1)\geq0,
\label{eq:flat-three-derivative-pi}
\end{align}
which is gauge invariant due to \eqref{eq:y_i^2y_i+1y_i-1 flat-space-schouten-identities}. No further new parity-even TT vertices arise at higher derivative order: candidates with four or more derivatives either vanish by Schouten identities or fail gauge invariance \cite{Mkrtchyan:2017ixk}.
\section{Metric-like cubic vertices in AdS$_3$}
\label{Metric-like cubic vertices in AdS_3}
\subsection{Commutation of AdS Covariant Derivatives}
\label{Commutation of AdS Covariant Derivatives}
The main hurdle in curved spacetime arises from the non-commutativity of covariant derivatives. In a maximally symmetric AdS background, this is encoded in the Riemann tensor
\begin{equation}
R_{\mu\nu\lambda\rho}
=
-\frac{1}{\ell^2}
\left(
g_{\mu\lambda}g_{\nu\rho}
-
g_{\mu\rho}g_{\nu\lambda}
\right),
\label{eq:riemann-tensor}
\end{equation}
where $\ell$ is the AdS radius. A commutator of derivatives acting directly on the field generating functions can be represented as
\begin{equation}
[{\nabla_i}_\mu, {\nabla_j}_\nu]\phi_1\phi_2\phi_3 = \delta_{ij}{a_i}_{\lambda}{\partial_i}^{\sigma}{R^{\lambda}}_{\sigma \mu\nu}\phi_1\phi_2\phi_3,\hspace{0.75cm}i,j\in \{1,2,3\},
\label{commutator operator form}
\end{equation}
where the $\partial_i$ acting on $\phi_i$ produces a sum contracting the Riemann tensor ${R^{\lambda}}_{\sigma\mu\nu}$ with each field index, while the additional explicit $a_{i\lambda}$ on the left restores the auxiliary vector with the appropriate index structure, see e.g. \cite{Kessel:2017mxa}. The $\delta_{ij}$ encodes the fact that covariant derivatives acting on different fields commute. We will also need the standard rule for expressions with additional spacetime indices on the right of the commutator:
\begin{align}
[{\nabla_i}_{\mu}, {\nabla_i}_{\nu}]
{\nabla_i}_{\rho_1}\cdots{\nabla_i}_{\rho_n}
\phi_1\phi_2\phi_3
&=
\Bigl(
{a_i}_{\lambda}{\partial_i}^{\sigma}{R^{\lambda}}_{\sigma\mu\nu}
{\nabla_i}_{\rho_1}\cdots{\nabla_i}_{\rho_n}
- {R^{\sigma}}_{\rho_1\mu\nu}
{\nabla_i}_{\sigma}\cdots{\nabla_i}_{\rho_n}\nonumber\\
&\quad
\hspace{0.7cm}-\cdots
- {R^{\sigma}}_{\rho_n\mu\nu}
{\nabla_i}_{\rho_1}\cdots{\nabla_i}_{\sigma}
\Bigr)
\phi_1\phi_2\phi_3 .
\end{align}
Lastly, any explicit \(\partial_i\) operators, including those contained in \(z_{i+1}\) or \(z_{i-1}\), should be moved to the left of the commutator before substituting \eqref{commutator operator form}. Schematically,
\begin{equation}
\begin{split}
[{\nabla_i}_\mu,{\nabla_i}_\nu]\cdot
\Bigl(\text{any } \partial_i,\, z_{i+1},\, z_{i-1}\Bigr)
\cdot \phi_1\phi_2\phi_3
\\[1mm]
=
\Bigl(\text{any } \partial_i,\, z_{i+1},\, z_{i-1}\Bigr)
\cdot
{a_i}_{\lambda}{\partial_i}^{\sigma}
{R^{\lambda}}_{\sigma\mu\nu}
\phi_1\phi_2\phi_3 .
\end{split}
\label{eq:commutator-ordering-rule}
\end{equation}
This ordering ensures that the curvature operator acts on the full set of auxiliary-vector indices carried by the \(i\)th field. If the \(\partial_i\)'s act first, they can prematurely remove auxiliary vectors \(a_i\), so that \eqref{commutator operator form} no longer reproduces the full set of curvature contractions with each field index.
\subsection{Generalised Gauge Variation Procedure}
\label{Generalised Gauge Variation Procedure}
In deriving the flat space gauge variation formula \eqref{differential gauge variation on general vertex}, we used the fact that the elementary commutators
\begin{equation}
[y_j, a_i\cdot\nabla_i],
\qquad
[z_j, a_i\cdot\nabla_i],
\label{general y,z commutators with anabla for example}
\end{equation}
themselves commute with the $y_k,z_k$ operators. This allows the terms generated by repeated use of the Leibniz rule to be straightforwardly collected into the action of a differential operator. For example,
\begin{multline}
[z_{i+1}^n, a_i\cdot\nabla_i]
=
z_{i+1}^{n-1}[z_{i+1}, a_i\cdot\nabla_i]
+
z_{i+1}^{n-2}[z_{i+1}, a_i\cdot\nabla_i]z_{i+1}
+\cdots
+
[z_{i+1}, a_i\cdot\nabla_i]z_{i+1}^{n-1}
\\
=
n z_{i+1}^{n-1}[z_{i+1}, a_i\cdot\nabla_i]
=
[z_{i+1}, a_i\cdot\nabla_i]\,
\partial_{z_{i+1}}(z_{i+1}^n).
\label{commutator for z power example for comparison to AdS}
\end{multline}
In AdS, however, the commutators \eqref{y and z commutators} become
\begin{align} 
[z_i, (a_i\cdot \nabla_i)] &= 0
\label{zi commutator with anabla in AdS}\\ 
[z_{i\pm1}, (a_i\cdot \nabla_i)] &= \partial_{i\mp1}\cdot \nabla_i
\label{zi+-1 commutator with anabla in AdS}\\
[y_i, (a_i\cdot \nabla_i)] &= \nabla_i\cdot \nabla_{i+1}
\label{yi commutator with anabla in AdS}\\
[y_{i+1}, (a_i\cdot \nabla_i)] &= 0
\label{yi+1 commutator with anabla in AdS}\\
[y_{i-1}, (a_i\cdot \nabla_i)] &= {a_{i}}_{\beta} {\partial_{i-1}}_{\alpha}[{\nabla_i}^{\alpha}, {\nabla_i}^{\beta}].
\label{yi-1 commutator with anabla in AdS}
\end{align}
This indicates that although \eqref{commutator for z power example for comparison to AdS} still holds for the $z_i$ operators, an analogous simplification with, say, $y_{i-1}$, whose commutator now contains non-commuting covariant derivatives, does not hold. Additionally, because in AdS the translation of these commutators into the $y_i,z_i$ basis depends on their initial ordering - whether they must be commuted past other operators for integration by parts or so that they can act directly on the fields - we have left the expressions in \eqref{zi commutator with anabla in AdS}-\eqref{yi-1 commutator with anabla in AdS} in terms of the original basic operators. Only once a specific operator placement is chosen can one unambiguously reduce them to the minimal $y_i,z_i$ basis. In practice, then, we need to formulate a systematic way to treat gauge variation, even though an explicit differential form for
\begin{equation}
[\mathcal{V}(y_i,z_i), (a_i\cdot\nabla_i)]
\end{equation}
is not readily available. One always begins by applying the Leibniz rule to obtain a sum of expressions containing sets of operators multiplying the fundamental commutators \eqref{zi commutator with anabla in AdS}-\eqref{yi-1 commutator with anabla in AdS}. Then we work to translate the expression into the minimal $y_i, z_i$ basis as follows:

\begin{itemize}
\item Whenever an explicit auxiliary vector $a_i$ arises, it is usually most convenient to immediately \textbf{commute it to the left of all $\partial_i$, $z_{i+1}$, and $z_{i-1}$}. This allows us to drop the resulting term, since all auxiliary vectors are set to zero in the final expression \eqref{making cubic action local}. It also allows us to freely rearrange auxiliary-vector derivatives $\partial_i$ before evaluating derivative commutators, which may otherwise introduce difficulties in \eqref{eq:commutator-ordering-rule}.

\item Terms of the form \(\partial_i\cdot \nabla_{i-1}\) are \textbf{commuted through to the left of all \(\nabla\)'s}, at which point partial integration allows us to replace them with 
\begin{equation}
\partial_i\cdot \nabla_{i-1} \mapsto -y_i-\operatorname{Div}_i .
\end{equation}

\item Likewise, terms of the form \(\nabla_i\cdot \nabla_{i+1}\) are \textbf{commuted to the left of all \(\nabla\)'s}, and partial integration allows us to replace them with
\begin{equation}
\nabla_i\cdot \nabla_{i+1} \mapsto \frac{1}{2}\left(\Box_{i-1} - \Box_i - \Box_{i+1}\right).
\end{equation}

\item \(\operatorname{Div}_i\) terms are \textbf{commuted to the right of all \(\nabla\)'s} to act directly on the fields, at which point they can be dropped due to the TT condition.

\item \(\Box_i\) terms are similarly \textbf{commuted to the right of all \(\nabla\)'s}, where we apply the on-shell conditions
\begin{equation}
\Box_i\phi_i \approx \frac{s_i(s_i-3)}{\ell^2}\phi_i
\label{field on-shell condition}
\end{equation}
for fields, and
\begin{equation}
\Box_i\xi_i \approx \frac{s_i(s_i-1)}{\ell^2}\xi_i
\label{gauge parameter on-shell condition}
\end{equation}
for gauge parameters. In applying the second relation, \(s_i\) is the spin of the field whose gauge parameter appears after variation. Here we have set \(d=3\) in the TT equations \eqref{AdS TT gauge on-shell conditions 1}, \eqref{AdS TT gauge parameter conditions 1}.

\item \(\operatorname{Tr}_i\) terms are \textbf{commuted to the right of all \(a_i\)'s}, at which point they can be dropped due to the TT condition.
\end{itemize}
\noindent The operator basis algorithm described above typically becomes intractable by hand, since repeated commutations rapidly increase the number of terms. We therefore implemented the algorithm in Python \cite{KingPomfretCode2026}, representing operator expressions as strings and applying the commutation, integration-by-parts, and on-shell rules automatically. A string-based implementation is convenient here because the calculation depends on the ordering of non-commuting operators and on position-dependent replacement rules. The steps listed above describe the algorithm used by the program. In the following sections, we mainly quote the final expressions, with representative intermediate steps and checks collected in the appendices. Further details for reproducing our calculations are provided in the accompanying code repository \cite{KingPomfretCode2026}.\\
\indent For the present work, we are interested only in the AdS extensions of the known two- and three-derivative flat space vertices. It is therefore sufficient to restrict $\mathcal{V}(y_i,z_i)$ to derivative dependence up to cubic order in the $y_i$, with the only cubic structure retained
being $y_1y_2y_3$. That is, the derivative dependence is restricted to terms involving $y_i$, $y_i y_j$, and $y_1y_2y_3$, while excluding cubic structures with repeated $y_i$'s, such as $y_i^2y_j$ and $y_i^3$. The $z_i$-dependence, however, can be kept general. In practice, one needs to compute
\begin{align}
[y_1^{n_1}y_2^{n_2}y_3^{n_3}z_1^{m_1}z_2^{m_2}z_3^{m_3}, (a_i\cdot\nabla_i)] \equiv [\mathcal{Y}(y)\mathcal{Z}(z), (a_i\cdot\nabla_i)]
\label{gauge variation of most general monomial}
\end{align}
with arbitrary powers $(n_i,m_i)$ so that the corresponding differential operator structure can be uniquely read off. During this calculation, we truncate to the sector of the gauge variation operator that acts non-trivially on the restricted vertex space specified above. In particular,
we retain terms up to the mixed cubic derivative
\(\partial_{y_1}\partial_{y_2}\partial_{y_3}\), but discard other cubic and higher \(y\)-derivative structures, such as
\(\partial_{y_i}^2\partial_{y_j}\) and \(\partial_{y_i}^3\), which do not contribute within this sector. Without loss of generality, one can set $i=1$ and restore cyclicity at the end. Applying the Leibniz rule to the factorised vertex monomial $\mathcal{Y}(y)\mathcal{Z}(z)$ and performing an integration by parts then yields
\begin{equation}
[\mathcal{Y}(y)\mathcal{Z}(z), \left(a_1\cdot\nabla_1\right)] = y_3\partial_{z_2}\mathcal{V} - y_2\partial_{z_3}\mathcal{V} - \partial_{2\mu}\nabla_2^{\mu}\partial_{z_3}\mathcal{V} + \mathcal{Z}(z)[\mathcal{Y}(y), \left(a_{1}\cdot\nabla_1\right)],
\label{initial aP commutator expansion}
\end{equation}
where $\mathcal{Y}(y)\mathcal{Z}(z)\mapsto\mathcal{V}$ has been used in the terms where no ambiguity arises. The first two terms reproduce the flat space gauge variation \eqref{differential gauge variation on general vertex}, while the last two encode the new AdS contributions. Putting the AdS corrections in reduced form using the algorithm described above results in
\begingroup
\allowdisplaybreaks[4]
\begin{align}
[\mathcal{V}(y,z),a_{1\mu}\nabla_1^\mu]
= \Bigg\{&
y_3\partial_{z_2}-y_2\partial_{z_3}
\nonumber\\
&+\frac{1}{2\ell^2}\Bigg[
z_1
\left(
y_3\partial_{y_3}
+2z_2\partial_{z_2}
-2z_3\partial_{z_3}
\right)
\partial_{y_3}\partial_{z_3}
\nonumber\\
&\qquad
+\Big(
s_3(s_3-3)-s_1(s_1-1)-s_2(s_2-3)
\nonumber\\
&\qquad\qquad
+2y_1z_1\partial_{y_2}\partial_{z_2}
-2(y_1z_1+y_2z_2)\partial_{y_3}\partial_{z_3}
\nonumber\\
&\qquad\qquad
-2y_2
\left(
1+z_1\partial_{z_1}-z_3\partial_{z_3}
\right)
\partial_{y_2}
\nonumber\\
&\qquad\qquad
+2z_3
\left(
2+z_1\partial_{z_1}+z_3\partial_{z_3}
\right)
\partial_{z_3}
\nonumber\\
&\qquad\qquad
+2y_3
\left(
1+z_3\partial_{z_3}
\right)
\partial_{y_3}
\Big)\partial_{y_1}
\nonumber\\
&\qquad
+\Big(
(y_2z_2-y_3z_3)\partial_{z_1}
+y_1\left(
2+3z_3\partial_{z_3}
\right)
\Big)\partial_{y_1}^2
\Bigg]
\displaybreak[3]\nonumber\\
&+\frac{1}{2\ell^4}z_1
\Bigg[
s_1(s_1-1)-s_2(s_2-3)+s_3(s_3-3)
\nonumber\\
&\qquad
+2z_2
\left(
1+z_1\partial_{z_1}
\right)
\partial_{z_2}
\nonumber\\
&\qquad
-2z_3
\left(
2+z_1\partial_{z_1}+2z_2\partial_{z_2}
\right)
\partial_{z_3}
\Bigg]
\partial_{y_1}\partial_{y_2}\partial_{y_3}
\nonumber\\
&+\mathcal{O}(\partial_{y_i}^2\partial_{y_j})
+\mathcal{O}(\partial_{y_i}^3)
\Bigg\}\mathcal{V}(y,z).
\label{AdS gauge variation only up to 2 (or 3) derivatives in y_i}
\end{align}
\endgroup
Further details of the intermediate steps are given in Appendix~\ref{app:ads-gauge-variation-diff-op}. We emphasise that \eqref{AdS gauge variation only up to 2 (or 3) derivatives in y_i} is a reduced formula tailored to the two- and three-derivative vertex sector considered in this work, rather than the general approach used internally by the Python implementation. In the Python code, the basis algorithm is performed at the level of the fully general ordered operator expressions described above, which is necessary for reducing the AdS dimension-dependent identities while also keeping the framework adaptable to possible higher-derivative extensions in future work. The formula \eqref{AdS gauge variation only up to 2 (or 3) derivatives in y_i} therefore provides an independent way to check the gauge variation of the vertices, and is implemented separately in Mathematica \cite{KingPomfretCode2026}.
\subsection{Dimension-dependent identities in AdS$_3$}
\label{Dimension-dependent identities in AdS_3}
The evaluation of the flat space Schouten identities in the $y_i,z_i$ basis in \cite{Mkrtchyan:2017ixk} is straightforward because the basic operators commute. Consequently, the same operator basis simplification can be applied independently of the surrounding operator structure, and any term proportional to the resulting identities vanishes identically. In AdS, this is no longer true. Covariant derivatives do not commute, so changing the ordering of derivatives, as noted in Section~\ref{Generalised Gauge Variation Procedure}, generates lower-derivative terms carrying powers of $1/\ell^2$. Thus the leading parts of the AdS$_3$ Schouten identities agree with their flat space counterparts, but they acquire curvature-dependent corrections. The precise form of these corrections depends on the surrounding operators in the original expression, since these determine which indices the derivative commutators act on and which explicit auxiliary vector structures are generated. 

It is important to note that, at the level of the bare over-antisymmetrisations, that is, the DDIs obtained by over-antisymmetrisation in isolation, there are no further independent DDI structures beyond the corresponding flat space expressions in \eqref{eq:(G - y_iz_i)^2 flat-space-schouten-identities}--\eqref{eq:y_i^2y_i+1y_i-1 flat-space-schouten-identities}. The only meaningful difference in AdS is the non-commutativity of the covariant derivatives. Thus, if one considers all possible bare over-antisymmetrisations, as in flat space, there are two possibilities. Either the leading flat space part is non-trivial, in which case one obtains the corresponding flat space DDI together with lower-derivative AdS corrections, or the leading flat space part vanishes identically, in which case only lower-derivative corrections remain. In the latter case, these lower-derivative corrections cannot define new independent bare DDI structures. Indeed, the bare DDI basis is obtained by selecting the independent expressions arising from all possible over-antisymmetrisations. Therefore, if an over-antisymmetrisation has a vanishing leading flat space part, any remaining lower-derivative correction must either coincide with one of the identities already selected in this basis, or vanish trivially. For example, four-derivative over-antisymmetrisations with vanishing leading components can leave two-derivative curvature corrections, which must already be included in the basis of two-derivative over-antisymmetrisations if they are non-trivial. By contrast, three- or two-derivative over-antisymmetrisations can only leave one- or zero-derivative corrections, which cannot arise from a non-trivial bare over-antisymmetrisation and hence vanish trivially. Bare over-antisymmetrisations with more than four derivatives either vanish trivially or reduce to total derivatives up to lower-derivative components, which, again, are already contained in the bare basis.\footnote{A more direct way to see that no further bare DDI structures can be generated in AdS is to note that the identities hold for arbitrary cosmological constant and can therefore be expanded in powers of $1/\ell^2$. Then, for any non-trivial over-antisymmetrisation, it is possible to choose an overall normalisation given by a suitable power of $\ell^2$ such that the flat space limit is non-trivial. This limit must correspond to a flat space DDI, meaning that there is a one-to-one map between the bare AdS DDI basis and the independent flat space DDI basis.}

To obtain a complete, independent set of Schouten identities in AdS, we must adopt a different approach. Rather than computing the antisymmetrisations in isolation, one must consider the more general operator environments in which each identity can appear. In the most general case, this takes the schematic form
\begin{equation}
[\text{antisymmetrised product of } \partial\text{'s and }\nabla\text{'s}]
y_1^{n_1}y_2^{n_2}y_3^{n_3}
z_1^{m_1}z_2^{m_2}z_3^{m_3}
\equiv 0.
\label{general AdS vertex for DDI}
\end{equation}
As was the case with gauge variation, the fully general case is highly non-trivial to compute and unnecessary for our purposes, so we make two useful observations. First, the simple commutation behaviour of the $z_i$ operators carries over from flat space, see \eqref{f(z) commutator}, allowing the $z_i$-dependence to be kept arbitrary. Second, since we are interested in AdS extensions of the flat space vertices \eqref{eq:flat-two-derivative-vertex} and \eqref{eq:flat-three-derivative-vertex}, it is sufficient to restrict attention to cases with
\begin{equation}
2 \leq 
n_1+n_2+n_3+N_y
\leq 4
\label{number of derivative constraint for flat space independent identity basis reduction}
\end{equation}
where $N_y$ is the number of derivatives present in the antisymmetrisation i.e. the number of $y_i$ in the corresponding flat space DDI. This is consistent with the flat space classification, where no higher $y$-degree DDIs are needed: multiplying the identities by additional powers of $y_i$ does not produce new information relevant for the two- and three-derivative vertex monomials considered in this work, because gauge variation \eqref{AdS gauge symmetry} can only produce terms with at most one more derivative than we started with. Even then, one need not consider every possible case satisfying \eqref{number of derivative constraint for flat space independent identity basis reduction}, because not all expressions obtained by multiplying lower derivative Schouten identities by additional $y_i$ are independent. This is best illustrated by example: consider a three-derivative expression obtained by multiplying the two-derivative Schouten identity \eqref{eq:(G - y_iz_i)^2 flat-space-schouten-identities} by $y_i$, 
\begin{align}
y_i \times (G - y_iz_i)^2 &= y_iy_{i+1}^2z_{i+1}^2 + y_iy_{i-1}^2z_{i-1}^2 + 2y_iy_{i-1}y_{i+1}z_{i-1}z_{i+1}
\nonumber\\ &= z_{i-1} \times y_iy_{i-1}\left(G-y_iz_i\right) + z_{i+1} \times y_iy_{i+1}\left(G-y_iz_i\right)
\label{(G-y_iz_i)^2 multiplied by y_i}
\end{align}
Thus, up to factors of $z_{i\pm1}$, this expression is a linear combination of the three-derivative identities \eqref{eq:yiyi-1(G - y_iz_i)flat-space-schouten-identities}, \eqref{eq:y_iy_i+1(G - y_iz_i) flat-space-schouten-identities}.
Since the $z_i$-dependence in \eqref{general AdS vertex for DDI} is kept arbitrary, the factors of $z_{i\pm1}$ in \eqref{(G-y_iz_i)^2 multiplied by y_i} can be absorbed into $\mathcal{V}(z)$. Therefore, the AdS extension of $y_i(G-y_iz_i)^2$ need not be computed separately: it is already contained in the AdS extensions of the three-derivative DDIs \eqref{eq:yiyi-1(G - y_iz_i)flat-space-schouten-identities}, \eqref{eq:y_iy_i+1(G - y_iz_i) flat-space-schouten-identities}.\\ 
\indent The preceding example suggests a systematic procedure for selecting an independent basis of Schouten identities in AdS. We first work in flat space, where operator ordering is not relevant, and consider products of each base Schouten identity \eqref{eq:(G - y_iz_i)^2 flat-space-schouten-identities}-\eqref{eq:y_i^2y_i+1y_i-1 flat-space-schouten-identities} with monomials of the form $y_1^{n_1}y_2^{n_2}y_3^{n_3}$ subject to \eqref{number of derivative constraint for flat space independent identity basis reduction}.
Among these products, we retain only an independent set modulo multiplication by arbitrary monomials in the \(z_i\). Once an independent flat space set has been selected, each representative is reverted to an operator expression of the form \eqref{general AdS vertex for DDI}. We then apply the AdS basis algorithm of the previous section to determine the curvature-dependent lower-derivative corrections generated by covariant-derivative commutators. One such independent set is:
\begin{align}
&(G-y_i z_i)^2,
\hspace{0.35cm}
y_i z_iG-y_{i+1}z_{i+1}y_{i-1}z_{i-1},
\label{flat space independent (y)(schouten identity) choice (1)}
\\
&y_i y_{i-1}(G-y_i z_i),
\hspace{0.35cm}
y_i y_{i+1}(G-y_i z_i),
\hspace{0.35cm}
y_{i-1}(G-y_i z_i)^2,
\\
&y_i^2y_{i+1}^2,
\hspace{0.35cm}
y_i^2y_{i+1}y_{i-1},
\hspace{0.35cm}
y_i y_{i-1}^2(G-y_i z_i),
\hspace{0.35cm}
y_i y_{i+1}^2(G-y_i z_i),
\hspace{0.35cm}
y_{i+1}^2(G-y_i z_i)^2 .
\label{flat space independent (y)(schouten identity) choice (2)}
\end{align}
Lifting these representatives to the operator environments in \eqref{general AdS vertex for DDI} yields an independent basis of AdS Schouten identities for the two- and three-derivative vertex sector considered in the present work. Their explicit forms, including the curvature-dependent lower-derivative terms, are recorded in full detail in Appendix~\ref{app:Complete Basis for DDIs}. These identities were checked in several ways. First, alongside the original computation in Python, an independent Mathematica implementation \cite{KingPomfretCode2026} was used to reproduce the steps taken to evaluate the DDIs. Second, one of the two-derivative DDIs was subjected to an explicit component-level check \cite{KingPomfretCode2026}. In this check, the trace, divergence, box, and integration-by-parts terms removed by the on-shell TT reduction were tracked explicitly and then systematically reintroduced. After expanding the antisymmetrised index components, this makes the vanishing of the original over-antisymmetrised expression manifest. The agreement among these checks gives strong evidence of the reliability of our results.
\subsection{Two- and three-derivative vertex corrections in AdS$_3$}
\label{Two- and three-derivative vertex corrections in AdS_3}
In the $\ell\rightarrow\infty$ limit, the AdS cubic vertices should reduce to the known flat space results. This suggests using ans\"atze of the form
\begin{equation}
\mathcal{V} = \mathcal{V}_0 + \frac{1}{\ell^2}\mathcal{V}_1 + \frac{1}{\ell^4}\mathcal{V}_2 + \cdots + \frac{1}{\ell^{2n}}\mathcal{V}_n,
\label{V = V_flat + V_AdS}
\end{equation}
where $\mathcal{V}_0$ is a known flat space vertex and the $\mathcal{V}_i$ with $1 \leq i \leq n$ represent AdS correction terms which are suppressed by factors of $\frac{1}{\ell^{2i}}$. Crucially, as seen in the evaluation of the Schouten identities, commutators of covariant derivatives will generate extra pieces with two fewer derivatives. In practice, we therefore build $\sum_{i=1}^n\frac{1}{\ell^{2i}}\mathcal{V}_i$ as a linear combination of every allowed operator monomial obtained by removing pairs of derivatives from $\mathcal{V}_0$. Each such term comes with an undetermined coefficient, which we then fix by demanding full gauge invariance. Note that $n$ in \eqref{V = V_flat + V_AdS} is then simply the maximum number of times one can remove a pair of derivatives from $\mathcal{V}_0$.
\subsubsection{Two-derivative AdS$_3$ vertex}
\label{Two-derivative AdS$_3$ vertex}
For the two-derivative vertex, removing a pair of $y_i$ leaves only $z_i$ dependence, whose structure is then fixed uniquely by
\begin{align}
\mathcal{V}_1
=
\alpha z_1^{n_1+1}z_2^{n_2+1}z_3^{n_3+1},
\label{two derivative AdS correction}
\end{align}
where
\begin{align}
n_i
=
\frac{1}{2}(s_{i-1}+s_{i+1}-s_i)-1
\geq 0,
\label{ni defs}
\end{align}
to have a complete contraction of all field indices, and $\alpha$ is to be fixed by gauge invariance. By cyclic symmetry, it suffices to impose gauge invariance under variation of the first field,
\begin{align}
[\mathcal{V}^{(2)}_{AdS},(a_1\cdot\nabla_1)]
\equiv{}&
[\mathcal{V}_0,(a_1\cdot\nabla_1)]
+\frac{1}{\ell^2}[\mathcal{V}_1,(a_1\cdot\nabla_1)]
\nonumber\\
={}&
\Big[
\left(
(s_3-1)y_3z_3
+(s_2-1)y_2z_2
+(s_1-1)y_1z_1
\right)
Gz_1^{n_1}z_2^{n_2}z_3^{n_3},
(a_1\cdot\nabla_1)
\Big]
\nonumber\\
&\quad
+\frac{\alpha}{\ell^2}
\Big[
z_1^{n_1+1}z_2^{n_2+1}z_3^{n_3+1},
(a_1\cdot\nabla_1)
\Big]
\equiv 0
\qquad (\mathrm{mod}\ \mathrm{DDIs}) .
\label{AdS gauge variation in first field two deriv vertex}
\end{align}
The gauge variation of \(\mathcal{V}_0\) contains the same three-derivative structures that vanish in flat space by virtue of the flat space Schouten identities. In AdS, eliminating these structures using the corresponding AdS identities generates additional one-derivative terms proportional to \(1/\ell^2\). These must be cancelled by the variation of the zero-derivative correction \(\mathcal{V}_1\), which also produces one-derivative terms. Reducing the variation to the \(y_i,z_i\) basis and applying \eqref{AdS DDI 3}, \eqref{AdS DDI 4} and \eqref{AdS DDI 5} to eliminate the three-derivative structures therefore leaves a set of one-derivative terms. Requiring their coefficients to vanish gives algebraic constraints on \(\alpha\), whose simultaneous solution is
\begin{align}
\alpha = \frac{1}{2}\Bigl(
  2 + s_1 + s_2 + s_3
  - 2s_1s_2 - 2s_1s_3 - 2s_2s_3
  + s_1s_2^2 + s_1s_3^2
\nonumber\\[-0.8ex]
  \quad
  +\,s_2s_1^2 + s_2s_3^2 + s_3s_1^2 + s_3s_2^2
  - 2s_1s_2s_3
  - s_1^3 - s_2^3 - s_3^3
\Bigr).
\label{alpha solution 2 deriv AdS vertex}
\end{align}
Intermediate details of this calculation are given in Appendix~\ref{Algebraic Constraints on Vertex Coefficients}. The coefficient is invariant under $i \mapsto i+1 \pmod{3}$ in the subscripts, as expected from the cyclic symmetry of the ansatz. To make contact with minimal gravitational coupling, it is useful to choose a DDI-equivalent representative that singles out the third field. Using
\eqref{AdS DDI 1}, the two-derivative vertex may be written as
\begin{align}
\mathcal{V}^{(2)}_{AdS}
\equiv{}&
y_3z_1^{n_1}z_2^{n_2}z_3^{n_3+1}
\Bigl[
(s_2+s_3-2)y_1z_1
+(s_3+s_1-2)y_2z_2
+(s_3-1)y_3z_3
\Bigr]
\nonumber\\
&+\frac{\alpha'}{\ell^2}
z_1^{n_1+1}z_2^{n_2+1}z_3^{n_3+1},
\label{minimal coupling form for AdS}
\end{align}
where
\begin{align}
\alpha'
= \frac{1}{2}\Bigl(&
6+s_1-3s_2-7s_3
-4s_1^2-4s_2^2+4s_3^2
+6s_1s_2+4s_2s_3
\nonumber\\
&\quad
+s_1^3+s_2^3-s_3^3
-s_1^2s_2-s_1s_2^2
-s_1s_3^2-s_2s_3^2
\nonumber\\
&\quad
+s_1^2s_3+s_2^2s_3
-2s_1s_2s_3
\Bigr).
\label{minimal coupling general s_1s_2s_3 coefficient}
\end{align}
This is no longer manifestly cyclically symmetric because the third
field has been singled out in anticipation of specialising it to a graviton, $s_3=2$. Doing this, while also setting $s_1=s_2=s$, gives
\begin{align}
\alpha'_{ss2}=-s(s+1).
\label{minimal coupling zero derivative coefficient}
\end{align}
The resulting vertex is the minimal gravitational coupling, which we verify
explicitly in Section~\ref{Consistency check for minimal coupling to gravity}.
\subsubsection{Three-derivative AdS$_3$ vertex}
\label{Three-derivative AdS$_3$ vertex}
Next, we find the lower derivative correction to the three-derivative vertex. Removing a pair of derivatives yields one-derivative expressions, so the most general AdS correction ansatz is
\begin{align}
\mathcal{V}_1 = \alpha_1y_1z_1^{p_1+1}z_2^{p_2}z_3^{p_3} + \alpha_2y_2z_1^{p_1}z_2^{p_2+1}z_3^{p_3} + \alpha_3y_3z_1^{p_1}z_2^{p_2}z_3^{p_3+1},
\label{AdS one derivative correction ansatz}
\end{align}
where 
\begin{align}
p_i = \frac{1}{2}\left(s_{i-1}+s_{i+1}-s_i-1\right).
\label{pi defs}
\end{align}
Just like the previous case, \eqref{V = V_flat + V_AdS} terminates at $\mathcal{V}_1$. Again, by cyclic symmetry, it suffices to impose gauge invariance under variation of just the first field:
\begin{align}
[\mathcal{V},(a_1\cdot\nabla_1)]
={}&
[\mathcal{V}_0,(a_1\cdot\nabla_1)]
+
\frac{1}{\ell^2}[\mathcal{V}_1,(a_1\cdot\nabla_1)]
\nonumber\\
={}&
\Big[
y_1y_2y_3z_1^{p_1}z_2^{p_2}z_3^{p_3},
(a_1\cdot\nabla_1)
\Big]
\nonumber\\
&+
\frac{1}{\ell^2}\Big[
\alpha_1y_1z_1^{p_1+1}z_2^{p_2}z_3^{p_3}
+\alpha_2y_2z_1^{p_1}z_2^{p_2+1}z_3^{p_3}
\nonumber\\
&\hspace{2.2cm}
+\alpha_3y_3z_1^{p_1}z_2^{p_2}z_3^{p_3+1},
(a_1\cdot\nabla_1)
\Big]
\equiv 0
\qquad (\mathrm{mod}\ \mathrm{DDIs}) .
\label{AdS gauge variation in first field three deriv vertex}
\end{align}
Now, the gauge variation of the flat part of the vertex, \(\mathcal{V}_0\), yields four-derivative terms which, in principle, can generate both two- and zero-derivative terms. Meanwhile, the one-derivative correction \(\mathcal{V}_1\) can produce two-derivative terms, and these may also generate zero-derivative terms. Therefore, after reducing \eqref{AdS gauge variation in first field three deriv vertex} to the \(y_i,z_i\) basis, using \eqref{AdS DDI 6} to eliminate the four-derivative terms, and subsequently applying \eqref{AdS DDI 1} to express the two-derivative terms in an independent basis free of $y_i^2$ structures, one finds a set of constraints among the remaining two- and zero-derivative terms, all of which are satisfied by
\begin{equation}
\alpha_i = \frac{1}{2}\left(2-2s_{i-1}-5s_i+s_is_{i+1}+s_is_{i-1}+s_i^2\right).
\qquad i\in\{1,2,3\}.
\label{alphai three derivative vertex 1 deriv coefficients}
\end{equation}
Again, further details of this calculation are recorded in Appendix~\ref{Algebraic Constraints on Vertex Coefficients}. The cyclic symmetry of the ansatz is reflected in the coefficients, as expected.
\subsection{Consistency check for minimal coupling to gravity}
\label{Consistency check for minimal coupling to gravity}
Minimal coupling to gravity may be extracted by expanding the covariantised Fronsdal action to first order in a spin-two perturbation of the AdS background metric, see e.g. \cite{Campoleoni:2012hp} for related work on the spin-3 case and its relation to the $SL(3,\mathbb{R})\times SL(3,\mathbb{R})$ Chern-Simons formulation. Here we use linearisation of the covariantised Fronsdal action as an independent consistency check of the AdS correction coefficient for the $s$-$s$-$2$ case of the two-derivative vertex given in \eqref{minimal coupling general s_1s_2s_3 coefficient}. As mentioned in Section~\ref{Two-derivative AdS$_3$ vertex}, DDIs allow the two-derivative vertex to be rewritten in the form \eqref{minimal coupling form for AdS}. In flat space, this form of the two-derivative vertex with
\begin{equation}
s_1=s_2=s,
\qquad
s_3=2,
\end{equation}
coincides with the \(s\text{-}s\text{-}2\) vertex generated by a minimal covariantisation of the Fronsdal action. To compute the curvature-dependent correction in AdS$_3$, we consider the part of the AdS$_3$ Fronsdal Lagrangian relevant in the TT sector,
\begin{align}
\mathcal{L}_{\mathrm{Frons.}}^{\mathrm{AdS}}
={}&
\sqrt{-g}\Big({\phi_2}^{\mu_1...\mu_s}\Box{\phi_1}_{\mu_1...\mu_s}
-\frac{s(s-3)}{\ell^2}
{\phi_2}^{\mu_1...\mu_s}{\phi_1}_{\mu_1...\mu_s}
\nonumber\\
&\quad
+\mathcal{O}\!\left((\nabla\cdot\phi)^2\right)
+\mathcal{O}\!\left((\operatorname{Tr}\phi)^2\right)
+\mathcal{O}\!\left(\operatorname{Tr}(\phi)\,(\nabla\cdot\phi)\right)\Big),
\label{initial linearised AdS fronsdal lagrangian}
\end{align}
and linearise with respect to a spin-two perturbation of the background metric $\eta_{\mu\nu}$,
\begin{equation}
g_{\mu\nu}= \eta_{\mu\nu}+h_{\mu\nu},
\qquad
g^{\mu\nu}= \eta^{\mu\nu}-h^{\mu\nu}+\mathcal{O}(h^2),
\label{metric linearisation and inverse}
\end{equation}
in order to pick out cubic $\mathcal{O}(\phi\phi h)$ terms in the reduced TT basis. The labels \(1\) and \(2\) on the spin-\(s\) fields are introduced to make contact with our cubic-vertex classification: the two entries are temporarily regarded as fields evaluated at distinct points, as in \eqref{making cubic action local}, so that the vertices are straightforwardly interpreted in terms of our non-local operators $y_i$ and $z_i$. The cubic vertices generated by expanding field contractions in \eqref{initial linearised AdS fronsdal lagrangian} are proportional to the free equations of motion and are therefore irrelevant due to field redefinition freedom. Expanding the metric contractions and the Christoffel symbol contributions from the covariant derivatives reproduces the two-derivative part of the \(s\text{-}s\text{-}2\) vertex. The non-commutativity of the AdS background derivatives additionally generates a curvature-dependent zero-derivative contribution. After reducing the resulting expression to the $y_i,z_i$ basis, the zero-derivative contribution is
\begin{equation}
\frac{1}{\ell^2}\mathcal{V}_1
=
\frac{s(s+1)}{\ell^2}\,
z_1z_2z_3^{s-1}.
\label{Fronsdal minimal coupling AdS correction}
\end{equation}
Together with the two-derivative contribution,
\begin{equation}
\mathcal{V}_0
=
-\left(
s y_1z_1+s y_2z_2+y_3z_3
\right)y_3z_3^{s-1},
\label{Fronsdal minimal coupling flat vertex}
\end{equation}
this is in agreement with \eqref{minimal coupling form for AdS} and \eqref{minimal coupling zero derivative coefficient}, up to an irrelevant overall sign. The explicit linearisation and term-by-term simplification are given in Appendix~\ref{Minimal Gravitational Coupling from the Fronsdal Action}.
\section{Summary and Conclusion}
\label{Summary and Conclusions}
In this work, we have constructed the AdS$_3$ extensions of the known parity-even two- and three-derivative metric-like cubic vertices for massless bosonic higher-spin fields in three dimensions. Starting from the flat space classification, we first developed a systematic procedure for evaluating gauge variations in AdS, where the non-commutativity of covariant derivatives generates curvature-dependent lower-derivative terms. We then derived the corresponding AdS corrections to the three-dimensional Schouten identities required for the vertex sector of interest. Using these ingredients, we imposed gauge invariance on the most general lower-derivative completions of the flat space vertices and fixed their coefficients.

For the two-derivative vertex, the AdS completion is obtained by adding a unique zero-derivative correction. For the three-derivative vertex, gauge invariance fixes the coefficients of the three possible one-derivative corrections. By construction, the resulting vertices reduce to the known flat space expressions in the limit \(\ell\to\infty\). As an independent check, we specialised the two-derivative result to the \(s\text{-}s\text{-}2\) coupling and recovered the curvature-dependent term obtained by expanding the AdS Fronsdal action to linear order in a spin-two metric perturbation.

Although we focused solely on bosonic vertices, a closely related flat space development is the recent classification of cubic interactions involving massless fermionic fields in three dimensions \cite{Fredenhagen:2024lps}. Extending the present AdS$_3$ construction to those fermionic vertices would therefore be a natural next step. A distinct, more ambitious extension concerns the inclusion of matter fields. Matter-coupled higher-spin gravities in three dimensions are strongly constrained in both flat and AdS$_3$ backgrounds \cite{Fredenhagen:2024kqn}, but identifying possible constructions with enlarged field content would be an interesting direction to pursue \cite{Sharapov:2024euk}.

There are also other immediate directions to take this work. Given the cubic action, the off-shell version of the $\mathcal{O}(g)$ condition from the Noether procedure \eqref{eq:cubic-vertex-gauge-symmetry} reads
\begin{align}
\delta_1^{\xi}S_2 + \delta_0^{\xi}S_3 = 0,
\label{off-shell O(g) Noether procedure condition}
\end{align}
leaving the first‐order deformation of the gauge symmetry $\delta_1^{\xi}$ as the only remaining unknown. By adapting the method of \cite{Joung:2013nma} to AdS$_3$, where we must systematically account for the Schouten identities as in \cite{Neckam:2023}, one can solve for $\delta_{1}^{\xi}$ explicitly. This requires knowledge of the off-shell Schouten identities, but these are easily obtained by replacing the $m_i$ in \eqref{AdS DDI 1}-\eqref{AdS DDI 10} with the corresponding d'Alembertians $\Box_i$ placed in the furthest right position in each term. This then enables analysis of the theory’s global symmetries, such as a direct check of the first-order closure of the gauge algebra and the evaluation of Jacobi identities for the gauge brackets, as performed in \cite{Neckam:2023} for spin-3 gravity in flat space.\\
\indent Another point of interest that has been left for future work is the construction of the fully general AdS analogue of \eqref{differential gauge variation on general vertex}, that is, obtaining a closed differential operator for gauge variation in AdS without restriction to a particular vertex sector. In principle, the commutators \eqref{[nabla,nabla] direct commutator}-\eqref{[f(y),ai]nabla expression} are all one needs to perform this calculation, though it is highly non-trivial in practice. Having such a closed‐form gauge variation operator would not only make general calculations more tractable, but would also provide a crucial foundation for exhaustively solving the AdS vertices without necessarily having knowledge of the flat space ones.

\section*{Acknowledgements}

K.M. is grateful to Stefan Fredenhagen for insightful discussions related to the subject of this work.

F.K. acknowledges support from the Science and Technology Facilities Council (STFC) through a PhD studentship during the preparation of this manuscript.

T.P. acknowledges support from the European Union through a PhD funded by the European Research Council (ERC) grant MW-ATLAS, grant number 101166905.

K.M. was partly supported by UKRI and STFC Consolidated Grant  ST/X000575/1 and by the Higher Education and Science Committee of the Republic of Armenia, under the Remote Laboratory Program, grant number 24RL-1C047. 

\appendix
\section{AdS Gauge variation differential operator up to $\partial_{y_1}\partial_{y_2}\partial_{y_3}$}
\label{app:ads-gauge-variation-diff-op}
To reduce the AdS corrections to the gauge variation operator in \eqref{initial aP commutator expansion}, one needs the commutator relations
\begingroup
\small
\allowdisplaybreaks
\begin{align}
[\nabla_{i\mu}, \nabla_{i\nu}]_{\mathrm{direct}}
=&\; \frac{1}{\ell^2}
\left(
a_{i\nu}\partial_{i\mu}
-
a_{i\mu}\partial_{i\nu}
\right),
\label{[nabla,nabla] direct commutator}\\
[\nabla_{i\mu}, \nabla_{i\nu}]\nabla_{\sigma}
=&\; [\nabla_{i\mu}, \nabla_{i\nu}]_{\mathrm{direct}}\nabla_{\sigma}
+
\frac{1}{\ell^2}
\left(
\delta_{\nu\sigma}\nabla_{\mu}
-
\delta_{\mu\sigma}\nabla_{\nu}
\right),
\\
[f(z), a_{i}^{\mu}]
=&\;
\left(
\partial_{i-1}^{\mu}\partial_{z_{i+1}}
+
\partial_{i+1}^{\mu}\partial_{z_{i-1}}
\right)f(z),
\label{f(z) commutator}
\\[1ex]
[f(y_i), \nabla_{i+1}^{\mu}]
=&\;
\colorbox{gray!15}{$
\frac{1}{\ell^2}
\left(
a_{i+1}^{\mu}z_{i-1}
-
a_{i+1}^{\nu}\partial_{i\nu}\partial_{i+1}^{\mu}
\right)\partial_{y_i}f(y_i)
+
\frac{1}{2\ell^2}
\Big(
\partial_i^{\mu}y_i\partial_{y_i}^2f(y_i)
$}
\nonumber\\
&\quad
-
\partial_{i\nu}\partial_i^{\nu}
\partial_{y_i}^2f(y_i)\nabla_{i+1}^{\mu}
\Big)
\nonumber\\
&\quad
+
\frac{1}{6\ell^4}
\left(
a_{i+1}^{\mu}z_{i-1}
-
a_{i+1}^{\sigma}\partial_{i\sigma}\partial_{i+1}^{\mu}
\right)
\partial_{i\nu}\partial_i^{\nu}
\partial_{y_i}^3f(y_i)
\nonumber\\
&\quad
+
\frac{1}{24\ell^2}
\partial_i^{\mu}\partial_{i\nu}\partial_i^{\nu}
y_i\partial_{y_i}^4f(y_i),
\label{[f(y),nabla] expressions form 1}
\\
=&\;
\colorbox{gray!15}{$
\frac{1}{\ell^2}
\left(
a_{i+1}^{\mu}z_{i-1}
-
a_{i+1}^{\nu}\partial_{i\nu}\partial_{i+1}^{\mu}
\right)\partial_{y_i}f(y_i)
+
\frac{1}{2\ell^2}
\Big(
\partial_i^{\mu}y_i\partial_{y_i}^2f(y_i)
$}
\nonumber\\
&\quad
-
\partial_{i\nu}\partial_i^{\nu}
\nabla_{i+1}^{\mu}\partial_{y_i}^2f(y_i)
\Big)
\nonumber\\
&\quad
-
\frac{1}{3\ell^4}
\left(
a_{i+1}^{\mu}z_{i-1}
-
a_{i+1}^{\sigma}\partial_{i\sigma}\partial_{i+1}^{\mu}
\right)
\partial_{i\nu}\partial_i^{\nu}
\partial_{y_i}^3f(y_i)
\nonumber\\
&\quad
-
\frac{5}{24\ell^2}
\partial_i^{\mu}\partial_{i\nu}\partial_i^{\nu}
y_i\partial_{y_i}^4f(y_i),
\label{[f(y),nabla] expressions form 2}
\\[1ex]
[f(y_i), a_i^{\mu}]
=&\;
\colorbox{gray!15}{$
\partial_{y_i}f(y_i)\nabla_{i+1}^{\mu}
-
\frac{1}{2\ell^2}
\left(
a_{i+1}^{\mu}z_{i-1}
-
a_{i+1}^{\nu}\partial_{i\nu}\partial_{i+1}^{\mu}
\right)\partial_{y_i}^2f(y_i)
-
\frac{1}{6\ell^2}
\partial_i^{\mu}y_i\partial_{y_i}^3f(y_i)
$}
\nonumber\\
&\quad
+
\frac{1}{6\ell^2}
\partial_{y_i}^3f(y_i)
\nabla_{i+1}^{\mu}\partial_{i\nu}\partial_i^{\nu}
\nonumber\\
&\quad
-
\frac{1}{24\ell^4}
\left(
a_{i+1}^{\mu}z_{i-1}
-
a_{i+1}^{\sigma}\partial_{i\sigma}\partial_{i+1}^{\mu}
\right)
\partial_{y_i}^4f(y_i)\partial_{i\nu}\partial_i^{\nu}
\nonumber\\
&\quad
-
\frac{1}{120\ell^4}
\partial_i^{\mu}y_i\partial_{y_i}^5f(y_i)
\partial_{i\nu}\partial_i^{\nu},
\label{[f(y),ai] expressions form 1}
\\
=&\;
\colorbox{gray!15}{$
\nabla_{i+1}^{\mu}\partial_{y_i}f(y_i)
+
\frac{1}{2\ell^2}
\left(
a_{i+1}^{\mu}z_{i-1}
-
a_{i+1}^{\nu}\partial_{i\nu}\partial_{i+1}^{\mu}
\right)\partial_{y_i}^2f(y_i)
+
\frac{1}{3\ell^2}
\partial_i^{\mu}y_i\partial_{y_i}^3f(y_i)
$}
\nonumber\\
&\quad
-
\frac{1}{3\ell^2}
\nabla_{i+1}^{\mu}\partial_{y_i}^3f(y_i)
\partial_{i\nu}\partial_i^{\nu}
\nonumber\\
&\quad
-
\frac{5}{24\ell^4}
\left(
a_{i+1}^{\mu}z_{i-1}
-
a_{i+1}^{\sigma}\partial_{i\sigma}\partial_{i+1}^{\mu}
\right)
\partial_{y_i}^4f(y_i)\partial_{i\nu}\partial_i^{\nu}
\nonumber\\
&\quad
-
\frac{2}{15\ell^4}
\partial_i^{\mu}y_i\partial_{y_i}^5f(y_i)
\partial_{i\nu}\partial_i^{\nu},
\label{[f(y),ai] expressions form 2}
\\[1ex]
[f(y_i), \nabla_{i+1}^{\mu}]\nabla_{i+1}^{\nu}
=&\;
\Bigg(
\frac{1}{\ell^2}
\left(
a_{i+1}^{\mu}z_{i-1}
-
a_{i+1}^{\sigma}\partial_{i\sigma}\partial_{i+1}^{\mu}
\right)\partial_{y_i}f(y_i)
+
\frac{1}{2\ell^2}
\partial_i^{\mu}y_i\partial_{y_i}^2f(y_i)
\Bigg)\nabla_{i+1}^{\nu}
\nonumber\\
&\quad
+
\frac{1}{\ell^2}
\partial_{y_i}f(y_i)
\left(
y_i\delta^{\mu\nu}
-
\partial_{i}^{\nu}\nabla_{i+1}^{\mu}
\right),
\label{[f(y),nabla]nabla expression}
\\
[f(y_i), a_i^{\mu}]\nabla_{i+1}^{\nu}
=&\;
\Bigg(
\partial_{y_i}f(y_i)\nabla_{i+1}^{\mu}
-
\frac{1}{2\ell^2}
\left(
a_{i+1}^{\mu}z_{i-1}
-
a_{i+1}^{\sigma}\partial_{i\sigma}\partial_{i+1}^{\mu}
\right)\partial_{y_i}^2f(y_i)
\nonumber\\
&\quad
-
\frac{1}{6\ell^2}
\partial_i^{\mu}y_i\partial_{y_i}^3f(y_i)
\Bigg)\nabla_{i+1}^{\nu}
-
\frac{1}{2\ell^2}
\partial_{y_i}^2f(y_i)
\left(
y_i\delta^{\mu\nu}
-
\partial_i^{\nu}\nabla_{i+1}^{\mu}
\right).
\label{[f(y),ai]nabla expression}
\end{align}
\endgroup
The `direct' subscript in \eqref{[nabla,nabla] direct commutator} indicates that the relation holds only when the commutator acts directly on the field generating functions. In equations \eqref{[f(y),nabla] expressions form 1}-\eqref{[f(y),ai] expressions form 2}, we have used the TT condition, assuming that at most one explicit auxiliary vector $a_i$ remains to the right of the commutator insertion. For these commutators, we display two algebraically equivalent forms corresponding to the two-derivative orderings that may arise in the calculation. For the specific cases considered in the following, only the highlighted leading terms are needed. The remaining terms would be required for constructing the gauge variation operator in full generality, which lies beyond the present scope. By contrast, equations \eqref{[f(y),nabla]nabla expression} and \eqref{[f(y),ai]nabla expression} assume that no explicit \(a_i\) remains to the right of the commutator, so that trace terms can be discarded immediately. As described in Section~\ref{Commutation of AdS Covariant Derivatives}, whenever an explicit auxiliary vector \(a_i\) is generated by a commutator, all \(\partial_i\), \(z_{i\pm1}\), and other \(a_i\) operators should first be moved to the left of the commutator insertion, while otherwise preserving relative orderings.\\
\indent It is useful to first analyse the final term in \eqref{initial aP commutator expansion}, as the necessary tools to treat the remaining correction will be contained within it. After writing $\mathcal{Y}(y)$ explicitly and inserting several instances of \eqref{[f(y),nabla] expressions form 1}-\eqref{[f(y),ai] expressions form 2}, while only retaining up to cubic $y$-derivatives of the form $\partial_{y_1}\partial_{y_2}\partial_{y_3}$, one finds
\begin{align}
[\mathcal{Y}(y),(a_1\cdot\nabla)]
=\Bigg[&
\frac{1}{\ell^2}a_{1\mu}a_1^{\mu}z_2\partial_{y_3}
+\frac{1}{\ell^2}a_{1\mu}\partial_3^{\mu}\partial_{y_3}
+\frac{1}{\ell^2}a_{1\mu}\nabla_2^{\mu}z_2
\partial_{y_1}\partial_{y_3}
\nonumber\\
&-\frac{1}{\ell^2}a_{1\mu}\partial_1^{\mu}a_{1\nu}\partial_3^{\nu}\partial_{y_3}
+\frac{1}{\ell^2}a_{1\mu}\partial_1^{\mu}y_3
\partial_{y_1}\partial_{y_3}
\nonumber\\
&-\frac{1}{\ell^4}a_{1\mu}\partial_1^{\mu}a_{3\nu}\partial_3^{\nu}z_1
\partial_{y_1}\partial_{y_2}\partial_{y_3}
-\frac{d-1}{\ell^4}a_{1\mu}\partial_1^{\mu}z_1
\partial_{y_1}\partial_{y_2}\partial_{y_3}
\nonumber\\
&+\frac{1}{2\ell^2}a_{1\mu}\partial_3^{\mu}y_3\partial_{y_3}^2
+\frac{1}{2}\left(\Box_3-\Box_1-\Box_2\right)\partial_{y_1}
\nonumber\\
&+\frac{1}{\ell^2}a_{1\mu}\nabla_2^{\mu}z_2
\partial_{y_1}\partial_{y_3}
-\frac{1}{\ell^2}a_{1\nu}\partial_3^{\nu}y_1
\partial_{y_1}\partial_{y_3}
\nonumber\\
&+\frac{1}{2\ell^2}a_{2\mu}\nabla_1^{\mu}z_3\partial_{y_1}^2
+\frac{1}{2\ell^2}a_{2\mu}\partial_1^{\mu}y_2\partial_{y_1}^2\nonumber\\
&+ \mathcal{O}(\partial_{y_i}^2\partial_{y_j}) + \mathcal{O}(\partial_{y_i}^3)
\Bigg]\mathcal{Y}(y).
\label{[V(y),anabla] first simplification}
\end{align}
To proceed, we first simplify the terms containing d'Alembertians, which, after commuting the pair of derivatives, can be shown to reduce to
\begin{align}
\Box_1\partial_{y_1}\mathcal{Y}(y)
=&\;
\Bigg[
m_1
-\frac{d-1}{\ell^2}y_3\partial_{y_3}
-\frac{2}{\ell^2}a_{1\mu}\nabla_1^\mu z_2\partial_{y_3}
\nonumber\\
&\qquad
+\mathcal{O}(\partial_{y_i}^2)
+\mathcal{O}(\partial_{y_1}\partial_{y_i})
+\mathcal{O}(\partial_{y_1}^2)
\Bigg]\partial_{y_1}\mathcal{Y}(y).
\label{box1 on partialy1 V(y)}
\end{align}

\begin{align}
\Box_2\partial_{y_1}\mathcal{Y}(y)
=&\;
\Bigg[
m_2
-\frac{d-1}{\ell^2}y_1\partial_{y_1}
-\frac{2}{\ell^2}a_{2\mu}\nabla_2^\mu z_3\partial_{y_1}
\nonumber\\
&\qquad
+\mathcal{O}(\partial_{y_i}^2)
+\mathcal{O}(\partial_{y_1}\partial_{y_i})
+\mathcal{O}(\partial_{y_1}^2)
\Bigg]\partial_{y_1}\mathcal{Y}(y).
\label{box2 on partialy1 V(y)}
\end{align}

\begin{align}
\Box_3\partial_{y_1}\mathcal{Y}(y)
=&\;
\Bigg[
m_3
-\frac{d-1}{\ell^2}y_2\partial_{y_2}
-\frac{2}{\ell^2}a_{3\mu}\nabla_3^\mu z_1\partial_{y_2}
\nonumber\\
&\qquad
-\frac{1}{\ell^2}(m_2-m_1-m_3)z_1
\partial_{y_2}\partial_{y_3}
\nonumber\\
&\qquad
+\mathcal{O}(\partial_{y_i}^2)
+\mathcal{O}(\partial_{y_1}\partial_{y_i})
+\mathcal{O}(\partial_{y_1}^2)
\Bigg]\partial_{y_1}\mathcal{Y}(y).
\label{box3 on partialy1 V(y)}
\end{align}
After substituting these relations into \eqref{[V(y),anabla] first simplification}, we use \eqref{f(z) commutator} to move the remaining auxiliary vectors past $\mathcal{Z}(z)$, where they can be dropped. This yields
\begin{align}
&\hspace{-1.1cm}\mathcal{Z}(z)[\mathcal{Y}(y), \left(a_{1}\cdot\nabla_1\right)]\nonumber\\
=\Bigg\{&
\frac{1}{\ell^2}z_1z_2\partial_{z_2}\partial_{z_3}\partial_{y_3}
-\frac{1}{\ell^2}z_1z_3\partial_{z_3}^2\partial_{y_3}
+\frac{1}{\ell^2}z_3y_3\partial_{z_3}\partial_{y_1}\partial_{y_3}
-\frac{1}{\ell^2}z_1y_1\partial_{z_3}\partial_{y_1}\partial_{y_3}
\nonumber\\
&-\frac{1}{2\ell^2}z_3y_3\partial_{z_1}\partial_{y_1}^2
+\frac{1}{2\ell^2}z_2y_2\partial_{z_1}\partial_{y_1}^2
-\frac{1}{\ell^2}z_1y_2\partial_{z_1}\partial_{y_1}\partial_{y_2}
+\frac{1}{2\ell^2}z_1y_3\partial_{z_3}\partial_{y_3}^2
\nonumber\\
&+\frac{1}{\ell^2}z_1y_1\partial_{z_2}\partial_{y_1}\partial_{y_2}
+\frac{1}{\ell^2}y_1z_3\partial_{z_3}\partial_{y_1}^2
-\frac{1}{\ell^2}z_2y_2\partial_{z_3}\partial_{y_1}\partial_{y_3}
+\frac{1}{2}(m_3-m_2-m_1)\partial_{y_1}
\nonumber\\
&-\frac{d-1}{2\ell^2}y_2\partial_{y_1}\partial_{y_2}
+\frac{d-1}{2\ell^2}y_1\partial_{y_1}^2
+\frac{d-1}{2\ell^2}y_3\partial_{y_1}\partial_{y_3}
\nonumber\\
&-\frac{1}{\ell^4}z_1\Big[
z_1z_2\partial_{z_1}\partial_{z_2}
+z_1z_3\partial_{z_1}\partial_{z_3}
+z_2z_3\partial_{z_2}\partial_{z_3}
+z_2^2\partial_{z_2}^2
+z_2\partial_{z_2}
\nonumber\\
&\hspace{3.4cm}
+(d-1)(z_2\partial_{z_2}+z_3\partial_{z_3})
+\frac{\ell^2}{2}(m_2-m_1-m_3)
\Big]\partial_{y_1}\partial_{y_2}\partial_{y_3}
\nonumber\\
&-\frac{2}{\ell^2}\partial_{3\mu}\nabla_3^{\mu}z_2\partial_{z_2}\partial_{y_1}\partial_{y_3}
+\frac{1}{\ell^2}\partial_{1\mu}\nabla_1^{\mu}z_1\partial_{z_2}\partial_{y_1}\partial_{y_2}+ \mathcal{O}(\partial_{y_i}^2\partial_{y_j}) + \mathcal{O}(\partial_{y_i}^3)
\Bigg\}\mathcal{V}(y_i,z_i),
\label{[V(y),anabla] boxes reduced + auxiliary vectors commuted}
\end{align}
Where the $m_i$ refer to the AdS masses defined in \eqref{field on-shell condition}-\eqref{gauge parameter on-shell condition}. It remains only to evaluate the final two terms, which contain divergences. Generally, one can show that
\begin{align}
\partial_{i\mu}\nabla_i^{\mu}\mathcal{V}(y_i,z_i)
=&\;
-\frac{1}{\ell^2}z_{i+1}
\Big[
z_{i+1}\partial_{z_{i+1}}
+z_{i-1}\partial_{z_{i-1}}
+y_i\partial_{y_i}
+\frac12 y_{i-1}\partial_{y_{i-1}}
+d-1
\nonumber\\
&\qquad
+\mathcal{O}(\partial_{y_j}^2)
+\mathcal{O}(\partial_{y_{i-1}}\partial_{y_j})
+\mathcal{O}(\partial_{y_{i-1}}^2)
\Big]
\partial_{y_{i-1}}\mathcal{V}(y_i,z_i).
\label{Div*V(y) reduction in y-deriv truncation}
\end{align}
from which it is straightforward to deduce
\begin{align}
&\partial_{3\mu}\nabla_3^{\mu}z_2\partial_{z_2}
\partial_{y_1}\partial_{y_3}\mathcal{V}(y_i,z_i)
\nonumber\\
&\qquad =
-\frac{1}{\ell^2}z_1z_2
\left(
z_1\partial_{z_1}
+z_2\partial_{z_2}
+d
\right)
\partial_{z_2}\partial_{y_1}\partial_{y_2}\partial_{y_3}
\mathcal{V}(y_i,z_i)
\nonumber\\
&\qquad\quad
+\mathcal{O}(\partial_{y_i}^2\partial_{y_j}\mathcal{V}(y_i,z_i))
+\mathcal{O}(\partial_{y_i}^3\mathcal{V}(y_i,z_i)).
\label{[V(y),anabla] div terms reduction U3nabla3}
\end{align}

\begin{align}
&\partial_{1\mu}\nabla_1^{\mu}z_1\partial_{z_2}
\partial_{y_1}\partial_{y_2}\mathcal{V}(y_i,z_i)
\nonumber\\
&\qquad =
-\frac{1}{\ell^2}z_1z_2
\left(
z_1\partial_{z_1}
+z_3\partial_{z_3}
+d-1
\right)
\partial_{z_2}\partial_{y_1}\partial_{y_2}\partial_{y_3}
\mathcal{V}(y_i,z_i)
\nonumber\\
&\qquad\quad
+\mathcal{O}(\partial_{y_i}^2\partial_{y_j}\mathcal{V}(y_i,z_i))
+\mathcal{O}(\partial_{y_i}^3\mathcal{V}(y_i,z_i)).
\label{[V(y),anabla] div terms reduction U1nabla1}
\end{align}
Equation \eqref{Div*V(y) reduction in y-deriv truncation} also gives us the other AdS correction term in \eqref{initial aP commutator expansion} for free:
\begin{align}
- \partial_{2\mu}\nabla_2^{\mu}\partial_{z_3}\mathcal{V}(y_i,z_i) = \frac{1}{\ell^2}z_3\left[z_3\partial_{z_3}+z_1\partial_{z_1}+d-1+y_2\partial_{y_2}+\frac12y_1\partial_{y_1}\right]\partial_{z_3}\partial_{y_1}\mathcal{V}(y_i,z_i).
\label{U2nabla2partialz3V reduction}
\end{align}
Putting everything together and simplifying, the final result is as follows:
\begingroup
\allowdisplaybreaks[4]
\begin{align}
[\mathcal{V}(y,z),a_{1\mu}\nabla_1^\mu]
= \Bigg\{&
y_3\partial_{z_2}-y_2\partial_{z_3}
\nonumber\\
&+\frac{1}{2\ell^2}\Bigg[
z_1
\left(
y_3\partial_{y_3}
+2z_2\partial_{z_2}
-2z_3\partial_{z_3}
\right)
\partial_{y_3}\partial_{z_3}
\nonumber\\
&\qquad
+\Big(
s_3(s_3-3)-s_1(s_1-1)-s_2(s_2-3)
\nonumber\\
&\qquad\qquad
+2y_1z_1\partial_{y_2}\partial_{z_2}
-2(y_1z_1+y_2z_2)\partial_{y_3}\partial_{z_3}
\nonumber\\
&\qquad\qquad
-y_2\left(
d-1
+2z_1\partial_{z_1}
-2z_3\partial_{z_3}
\right)\partial_{y_2}
\nonumber\\
&\qquad\qquad
+2z_3\left(
d-1
+z_1\partial_{z_1}
+z_3\partial_{z_3}
\right)\partial_{z_3}
\nonumber\\
&\qquad\qquad
+y_3\left(
d-1
+2z_3\partial_{z_3}
\right)\partial_{y_3}
\Big)\partial_{y_1}
\nonumber\\
&\qquad
+\Big(
(y_2z_2-y_3z_3)\partial_{z_1}
+y_1\left(
d-1
+3z_3\partial_{z_3}
\right)
\Big)\partial_{y_1}^2
\Bigg]
\displaybreak[3]\nonumber\\
&+\frac{1}{2\ell^4}z_1
\Bigg[
s_1(s_1-1)-s_2(s_2-3)+s_3(s_3-3)
\nonumber\\
&\qquad
+2z_2\left(
1+z_1\partial_{z_1}
\right)\partial_{z_2}
\nonumber\\
&\qquad
-2z_3\left(
d-1
+z_1\partial_{z_1}
+2z_2\partial_{z_2}
\right)\partial_{z_3}
\Bigg]
\partial_{y_1}\partial_{y_2}\partial_{y_3}
\nonumber\\
&+\mathcal{O}(\partial_{y_i}^2\partial_{y_j})
+\mathcal{O}(\partial_{y_i}^3)
\Bigg\}\mathcal{V}(y,z).
\label{AdS gauge variation operator general d}
\end{align}
\endgroup
Setting $d=3$ results in \eqref{AdS gauge variation only up to 2 (or 3) derivatives in y_i} as given in the main text.
\section{Independent AdS$_3$ DDI Basis for the $y$-Truncated Vertex Sector}
\label{app:Complete Basis for DDIs}
Here, we list each Schouten identity obtained via the procedure described in Section~\ref{Dimension-dependent identities in AdS_3}, as well as the operator antisymmetrisations from which they are derived:
\begin{align}
&\nabla_{i+1[a}\partial_{ib}\partial_{i+1c}\partial_{i-1d]}\nabla_{i-1}^a\partial_i^b\partial_{i+1}^c\partial_{i-1}^d\mathcal{V}(z) \equiv 0 \;\;\Rightarrow\nonumber\\
\nonumber\\
&\left(y_iz_iG-y_{i-1}z_{i-1}y_{i+1}z_{i+1}\right)\mathcal{V}(z)\nonumber\\ &\equiv z_1z_2z_3\left[m_i-m_{i+1}-m_{i-1}+\frac{1}{\ell^2}\left(z_{i-1}\partial_{z_{i-1}}-z_{i+1}\partial_{z_{i+1}}-2z_i\partial_{z_i}-3\right)\right]\mathcal{V}(z)
\label{AdS DDI 1}
\end{align}
\begin{align}
\nonumber\\
&\nabla_{i[a}\partial_{ib}\partial_{i+1c}\partial_{i-1d]}\nabla_i^a\partial_i^b\partial_{i+1}^c\partial_{i-1}^d\mathcal{V}(z) \equiv 0 \;\;\Rightarrow\nonumber\\
\nonumber\\
&(G-y_iz_i)^2\mathcal{V}(z) \equiv -z_1z_2z_3\left[2m_i+ \frac{1}{\ell^2}\left(3z_{i-1}\partial_{z_{i-1}}+z_{i+1}\partial_{z_{i+1}}+6\right)\right]\mathcal{V}(z)
\label{AdS DDI 2}
\end{align}
\begin{align}
&\nabla_{i[a}\nabla_{i+1b}\partial_{ic}\partial_{i-1d]}\nabla_i^a\partial_i^b\partial_{i+1}^c\partial_{i-1}^d \equiv 0\;\;\Rightarrow\nonumber\\
\nonumber\\
&y_iy_{i-1}(G-y_iz_i)\mathcal{V}(z)\nonumber\\&
\equiv
\left\{\,
\begin{aligned}
  &y_{i-1}z_{i-1}z_{i+1}\Bigl[\tfrac12\bigl(m_i+m_{i-1}-m_{i+1}\bigr)
    +\tfrac1{\ell^2}\bigl(
        2z_{i-1}\partial_{z_{i-1}}
      + z_{i+1}\partial_{z_{i+1}}
      + z_i\partial_{z_i}
      + 6
      \bigr)
  \Bigr]\\
  &\quad+\,\tfrac12\bigl(m_{i-1}-m_i-m_{i+1}\bigr)\,y_{i+1}z_{i+1}^2
  -\,y_i z_i z_{i+1}\Bigl[m_i
    + \tfrac{2}{\ell^2}\bigl(z_{i-1}\partial_{z_{i-1}}+1\bigr)
  \Bigr]
\end{aligned}
\right\}\;\mathcal V(z)
\label{AdS DDI 3}
\end{align}
\begin{align}
&\nabla_{i-1[a}\nabla_{ib}\partial_{ic}\partial_{i+1d]}\nabla_i^a\partial_{i-1}^b\partial_i^c\partial_{i+1}^d\mathcal{V}(z) \equiv 0\;\;\Rightarrow\nonumber\\
\nonumber\\
&y_iy_{i+1}(G-y_iz_i)\mathcal{V}(z)\nonumber\\
&\equiv
\left\{\,
\begin{aligned}
  &y_{i+1}z_{i+1}z_{i-1}\left[\frac{1}{2}\left(m_{i+1}+m_i-m_{i-1}\right) + \frac{2}{\ell^2}\left(z_i\partial_{z_i}+z_{i-1}\partial_{z_{i-1}}+3\right)\right]\\
  &+y_{i-1}z_{i-1}^2\left[\frac{1}{2}\left(m_{i+1}-m_i-m_{i-1}\right)+\frac{1}{\ell^2}\left(z_i\partial_{z_i}-z_{i+1}\partial_{z_{i+1}}\right)\right]\\
  &-y_iz_iz_{i-1}\left[m_i+\frac{1}{\ell^2}\left(z_{i+1}\partial_{z_{i+1}}+z_{i-1}\partial_{z_{i-1}}+3\right)\right]
\end{aligned}
\right\}\;\mathcal V(z)
\label{AdS DDI 4}
\end{align}
\begin{align}
&\nabla_{i[a}\partial_{ib}\partial_{i+1c}\partial_{i-1d]}\nabla_i^a\partial_i^b\partial_{i+1}^c\partial_{i-1}^dy_{i-1}\mathcal{V}(z)\equiv 0\;\;\Rightarrow\nonumber\\
\nonumber\\
&y_{i-1}(G-y_iz_i)^2\mathcal{V}(z) \equiv
\left\{\,
\begin{aligned}
  y_{i-1}z_1z_2z_3\left[-2m_i+\frac{1}{\ell^2}\left(z_{i+1}\partial_{z_{i+1}}-5z_{i-1}\partial_{z_{i-1}}-4\right)\right]\nonumber\\
  +\frac{1}{\ell^2}y_{i+1}z_{i+1}^2z_i\left(2z_{i+1}\partial_{z_{i+1}}-2z_{i-1}\partial_{z_{i-1}}+3\right)
\end{aligned}
\right\}\;\mathcal V(z)\\
\label{AdS DDI 5}
\end{align}
\begin{align}
&\nabla_{i-1[a}\nabla_{ib}\partial_{ic}\partial_{i-1d]}\nabla_{i-1}^a\nabla_i^b\partial_i^c\partial_{i+1}^d \mathcal{V}(z) \equiv 0\;\;\Rightarrow\nonumber\\
\nonumber\\
&y_i^2y_{i-1}y_{i+1}\mathcal{V}(z)\nonumber\\
&\equiv
\left\{\,
\begin{aligned}
  &y_iy_{i-1}z_{i-1}\left[\frac{1}{2}\left(m_{i+1}-m_{i-1}-m_i\right)+\frac{1}{\ell^2}\left(2z_i\partial_{z_i}-z_{i+1}\partial_{z_{i+1}}\right)\right]\nonumber\\
  &y_iy_{i+1}z_{i+1}\left[\frac{1}{2}\left(m_{i-1}-m_{i+1}-m_i\right)+\frac{1}{\ell^2}\left(z_{i-1}\partial_{z_{i-1}}+1\right)\right]+\frac{1}{\ell^2}y_{i+1}y_{i-1}z_{i+1}z_{i-1}\partial_{z_i}\nonumber\\
  &-y_i^2z_i\left[m_i+\frac{1}{\ell^2}\left(z_{i-1}\partial_{i-1}+2\right)\right] + \frac{1}{\ell^2}y_{i+1}^2z_{i+1}^2\partial_{z_i}+\frac{1}{\ell^2}y_{i-1}^2z_{i-1}^2\partial_{z_i}\nonumber\\
  &z_{i+1}z_{i-1}\Bigl[\frac{1}{4}\left(m_i^2+m_{i+1}^2+m_{i-1}^2\right)-\frac{1}{2}\left(m_im_{i+1}+m_im_{i-1}+m_{i-1}m_{i+1}\right)\nonumber\\
  &+\frac{1}{2\ell^2}\left(m_i-3m_{i-1}-5m_{i+1}\right)+\frac{1}{2\ell^2}\left(m_i-m_{i+1}-m_{i-1}\right)z_{i-1}\partial_{z_{i-1}}\nonumber\\
  &-\frac{1}{2\ell^2}\left(m_{i-1}-m_{i+1}-m_i\right)z_i\partial_{z_i}-\frac{1}{\ell^2}m_{i+1}z_{i+1}\partial_{z_{i+1}}\nonumber\\
  &-\frac{1}{\ell^4}\left(z_i\partial_{z_i}+z_{i+1}\partial_{z_{i+1}}+2z_{i-1}\partial_{z_{i-1}}+5\right)\Bigr]
\end{aligned}
\right\}\;\mathcal V(z)\\
\label{AdS DDI 6}
\end{align}
\begin{align}
&\nabla_{i[a}\nabla_{i+1b}\partial_{i+1c}\partial_{id]}\nabla_{i+1}^a\nabla_{i-1}^b\partial_i^c\partial_{i+1}^d \mathcal{V}(z) \equiv 0\;\;\Rightarrow\nonumber\\
\nonumber\\
&y_i^2y_{i+1}^2\mathcal{V}(z)\nonumber\\
&\equiv
\left\{\,
\begin{aligned}
  &y_iy_{i+1}z_{i-1}\Bigl[m_i+m_{i+1}-m_{i-1}+\frac{2}{\ell^2}\left(2z_i\partial_{z_i}+z_{i-1}\partial_{z_{i-1}}+5\right)\Bigr]\nonumber\\
  &-\frac{1}{\ell^2}y_i^2z_iz_{i-1}\partial_{z_{i+1}} -\frac{1}{\ell^2}y_{i+1}^2z_{i+1}z_{i-1}\partial_{z_i}\nonumber\\
  &+z_{i-1}^2\Bigl\{-\frac{1}{4}\left(m_i^2+m_{i- 1}^2+m_{i+1}^2\right)
  +\frac{1}{2}\left(m_im_{i-1}+m_im_{i+1}+m_{i-1}m_{i+1}\right)\nonumber\\
  &+\frac{1}{\ell^2}\Bigl[\left(m_{i-1}-m_{i+1}\right)z_i\partial_{z_i}+m_{i+1}z_{i+1}\partial_{z_{i+1}}+m_{i-1}z_{i-1}\partial_{z_{i-1}}+4m_{i-1}\Bigr]\nonumber\\
  &+\frac{1}{\ell^4}\left(4z_{i+1}\partial_{z_{i+1}}-7z_i\partial_{z_i}-2z_i^2\partial_{z_i}^2-z_iz_{i-1}\partial_{z_i}\partial_{z_{i-1}}+z_{i+1}z_{i-1}\partial_{z_{i+1}}\partial_{z_{i-1}}\right)\Bigr\}
\end{aligned}
\right\}\;\mathcal V(z)\\
\label{AdS DDI 7}
\end{align}
\begin{align}
&\nabla_{i[a}\nabla_{i+1b}\partial_{ic}\partial_{i-1d]}\nabla_i^a\partial_i^b\partial_{i+1}^c\partial_{i-1}^d y_{i-1}\mathcal{V}(z) \equiv 0\;\;\Rightarrow\nonumber\\
\nonumber\\
&y_iy_{i-1}^2(G-y_iz_i)\mathcal{V}(z)\nonumber\\
&\equiv
\left\{\,
\begin{aligned}
  &y_{i-1}y_{i+1}z_{i+1}^2\Bigl[\frac{1}{2}\left(m_{i-1}-m_{i+1}-m_i\right)+\frac{1}{\ell^2}\Bigr]-\frac{2}{\ell^2}y_iy_{i+1}z_iz_{i+1}^2\partial_{z_{i-1}}\nonumber\\
  &+y_iy_{i-1}z_iz_{i+1}\Bigl[-m_i+\frac{1}{\ell^2}\left(z_{i+1}\partial_{z_{i+1}}-3z_{i-1}\partial_{z_{i-1}}-1\right)\Bigr]-\frac{1}{\ell^2}y_{i+1}^2z_{i+1}^3\partial_{z_{i-1}}\nonumber\\
  &+y_{i-1}^2z_{i+1}z_{i-1}\Bigl[\frac{1}{2}\left(m_i+m_{i-1}-m_{i+1}\right)+\frac{1}{\ell^2}\left(3z_{i-1}\partial_{z_{i-1}}+z_{i+1}\partial_{z_{i+1}}+z_i\partial_{z_i}+8\right)\Bigr]\nonumber\\
  &+\frac{1}{\ell^2}z_iz_{i+1}^2\Bigl[\frac{1}{2}\left(m_{i+1}-m_{i-1}+3m_i\right)+\frac{1}{2}\left(m_i+m_{i+1}-m_{i-1}\right)\left(z_{i+1}\partial_{z_{i+1}}+z_{i-1}\partial_{z_{i-1}}\right)\nonumber\\
  &+\frac{1}{\ell^4}\left(3z_{i-1}^2\partial_{z_{i-1}}^2+7z_{i-1}\partial_{z_{i-1}}-z_{i-1}z_{i+1}\partial_{z_{i-1}}\partial_{z_{i+1}}+2z_iz_{i-1}\partial_{z_i}\partial_{z_{i-1}}+1\right)\Bigr]
\end{aligned}
\right\}\;\mathcal V(z)\\
\label{AdS DDI 8}
\end{align}
\begin{align}
&\nabla_{i-1[a}\nabla_{ib}\partial_{ic}\partial_{i+1d]}\nabla_i^a\partial_{i-1}^b\partial_i^c\partial_{i+1}^d  y_{i+1}\mathcal{V}(z) \equiv 0\;\;\Rightarrow\nonumber\\
\nonumber\\
&y_iy_{i+1}^2(G-y_iz_i)\mathcal{V}(z)\nonumber\\
&\equiv
\left\{\,
\begin{aligned}
  &y_{i-1}y_{i+1}z_{i-1}^2\Bigl[\frac{1}{2}\left(m_{i+1}-m_i-m_{i-1}\right)+\frac{1}{\ell^2}\left(-z_{i+1}\partial_{z_{i+1}}+2z_i\partial_{z_i}+2\right)\Bigr]\nonumber\\
  &-y_iy_{i+1}z_{i-1}z_i\Bigl[m_i+\frac{1}{\ell^2}\left(2z_{i+1}\partial_{z_{i+1}}+z_{i-1}\partial_{z_{i-1}}+4\right)\Bigr]-\frac{1}{\ell^2}y_iy_{i-1}z_{i-1}^2z_i\partial_{z_{i+1}}\nonumber\\
  &+y_{i+1}^2z_{i-1}z_{i+1}\Bigl[\frac{1}{2}\left(m_i+m_{i+1}-m_{i-1}\right)+\frac{1}{\ell^2}\left(3z_i\partial_{z_i}+2z_{i-1}\partial_{z_{i-1}}+9\right)\Bigr]\nonumber\\
  &+z_{i-1}^2z_i\Bigl[\frac{1}{2\ell^2}\left(m_{i+1}+3m_i-m_{i-1}\right)+\frac{m_i}{\ell^2}\left(z_i\partial_{z_i}+z_{i+1}\partial_{z_{i+1}}\right)\nonumber\\
  &+\frac{1}{\ell^4}\Bigl(z_{i+1}^2\partial_{z_{i+1}}^2+9z_{i+1}\partial_{z_{i+1}}+2z_iz_{i+1}\partial_{z_i}\partial_{z_{i+1}}+4z_i\partial_{z_i}\nonumber\\
  &+2z_{i-1}z_{i+1}\partial_{z_{i-1}}\partial_{z_{i+1}}+z_{i-1}z_i\partial_{z_{i-1}}\partial_{z_i}+2z_{i-1}\partial_{z_{i-1}}+7\Bigr)\Bigr]
\end{aligned}
\right\}\;\mathcal V(z)\\
\label{AdS DDI 9}
\end{align}
\begin{align}
&\nabla_{i[a}\partial_{ib}\partial_{i+1c}\partial_{i-1d]}\nabla_i^a\partial_i^b\partial_{i+1}^c\partial_{i-1}^d   y_{i+1}^2\mathcal{V}(z) \equiv 0\;\;\Rightarrow\nonumber\\
\nonumber\\
&y_{i+1}^2(G-y_iz_i)^2\mathcal{V}(z) \equiv -y_{i+1}^2z_1z_2z_3\left[2m_i+\frac{1}{\ell^2}\left(3z_{i-1}\partial_{z_{i-1}}+z_{i+1}\partial_{z_{i+1}}+6\right)\right]\mathcal{V}(z)
\label{AdS DDI 10}
\end{align}
Here, the AdS mass parameters $m_i$, $m_{i\pm1}$ have been left unevaluated so that the expressions remain applicable irrespective of which field is gauge varied. This basis is not unique; a more exhaustive analysis of \eqref{flat space independent (y)(schouten identity) choice (1)}-\eqref{flat space independent (y)(schouten identity) choice (2)} may admit a simpler choice of independent representatives.
\section{Algebraic Constraints on Vertex Coefficients}
\label{Algebraic Constraints on Vertex Coefficients}
Here we state the explicit forms of \eqref{AdS gauge variation in first field two deriv vertex} and \eqref{AdS gauge variation in first field three deriv vertex} after converting to the $y_i ,z_i$ basis via the process described in Section~\ref{Generalised Gauge Variation Procedure}, followed by their respective explicit forms after using DDIs to remove higher derivative terms. Without loss of generality, the calculations shown here follow from gauge variation with respect to the first field.\\
\indent Starting with the two-derivative vertex \eqref{AdS gauge variation in first field two deriv vertex}, expanding out the commutators and applying the AdS basis algorithm yields
\begingroup
\small
\allowdisplaybreaks[4]
\begin{align}
&\Bigl(
(s_1-1)y_1z_1\Bigl(n_2y_3z_3-n_3y_2z_2\Bigr)
+(s_2-1)y_2z_2\Bigl((n_2+1)y_3z_3-n_3y_2z_2\Bigr)
\nonumber\\
&\hspace{1.1cm}
+(s_3-1)y_3z_3\Bigl(n_2y_3z_3-(n_3+1)y_2z_2\Bigr)
\Bigr)
z_1^{n_1}z_2^{n_2-1}z_3^{n_3-1}G
\nonumber\\[1mm]
&\quad
+\frac{1}{\ell^2}\Bigg\{
y_1 z_1^{n_1+2}z_2^{n_2}z_3^{n_3}
\bigg(
3-\frac{13}{2}s_1+\frac{9}{2}s_1^2-s_1^3
-\frac{3}{2}s_2+3s_1s_2+\frac{1}{2}s_1^2s_2
\nonumber\\
&\hspace{3.6cm}
+\frac{1}{2}s_2^2-\frac{1}{2}s_1s_2^2
+\frac{7}{2}s_3-6s_1s_3+\frac{1}{2}s_1^2s_3
-s_2s_3-\frac{1}{2}s_1s_2s_3
\nonumber\\
&\hspace{3.6cm}
-\frac{1}{2}s_2^2s_3+\frac{1}{2}s_3^2
+s_1s_3^2+s_2s_3^2-\frac{1}{2}s_3^3
\bigg)
\nonumber\\[1mm]
&\qquad
+y_2 z_1^{n_1+1}z_2^{n_2+1}z_3^{n_3}
\bigg(
3-\frac{7}{2}s_1+\frac{3}{2}s_1^2-\frac{1}{2}s_1^3
-\frac{7}{2}s_2+\frac{5}{2}s_1s_2
\nonumber\\
&\hspace{3.6cm}
+2s_2^2-\frac{1}{2}s_2^3
+\frac{9}{2}s_3-\frac{5}{2}s_1s_3-4s_2s_3
-\frac{1}{2}s_1s_2s_3
\nonumber\\
&\hspace{3.6cm}
+s_1s_3^2+s_2s_3^2-\frac{1}{2}s_3^3
+\frac{\alpha}{2}\left(-s_1-s_2+s_3\right)
\bigg)
\nonumber\\[1mm]
&\qquad
+y_3 z_1^{n_1+1}z_2^{n_2}z_3^{n_3+1}
\bigg(
-3s_1+3s_1^2-\frac{1}{2}s_1^3
-2s_2+2s_1s_2+\frac{1}{2}s_1^2s_2
\nonumber\\
&\hspace{3.6cm}
+s_2^2+3s_3-3s_1s_3-\frac{1}{2}s_1^2s_3
-s_2s_3-s_1s_2s_3-s_2^2s_3
\nonumber\\
&\hspace{3.6cm}
+\frac{3}{2}s_1s_3^2+\frac{3}{2}s_2s_3^2-\frac{1}{2}s_3^3
+\frac{\alpha}{2}\left(s_1-s_2+s_3\right)
\bigg)
\Bigg\}.
\label{three derivative part of two deriv AdS gauge variation}
\end{align}
\endgroup
Then, the Schouten identities \eqref{AdS DDI 3}, \eqref{AdS DDI 4}, \eqref{AdS DDI 5} can be used to exchange the three-derivative part for one-derivative terms. Generally speaking, one can use these DDIs to exchange any term containing three $y_i$ with $y_1y_2y_3$ plus curvature dependent one-derivative terms. In \eqref{three derivative part of two deriv AdS gauge variation}, this causes the part containing $y_1$ to automatically vanish, while the remaining $y_2$ and $y_3$ parts impose two constraints on $\alpha$,
\begin{align}
&s_1+\frac{1}{2}s_1^2-\frac{1}{2}s_1^4
+s_2+s_1s_2-s_1^2s_2
+\frac{1}{2}s_2^2-s_1s_2^2+s_1^2s_2^2-\frac{1}{2}s_2^4
\nonumber\\
&\quad
-s_3-s_1^2s_3+s_1^3s_3
-s_1s_2s_3-s_1^2s_2s_3
-s_2^2s_3-s_1s_2^2s_3+s_2^3s_3
\nonumber\\
&\quad
-\frac{1}{2}s_3^2+s_1s_3^2+s_2s_3^2
+2s_1s_2s_3^2-s_1s_3^3-s_2s_3^3+\frac{1}{2}s_3^4
\nonumber\\
&\quad
+\alpha\left(-s_1-s_2+s_3\right)
=0,
\end{align}
\begin{align}
&-s_1-\frac{1}{2}s_1^2+\frac{1}{2}s_1^4
+s_2+s_1^2s_2-s_1^3s_2
+\frac{1}{2}s_2^2-s_1s_2^2+s_1s_2^3-\frac{1}{2}s_2^4
\nonumber\\
&\quad
-s_3-s_1s_3+s_1^2s_3
+s_1s_2s_3+s_1^2s_2s_3
-s_2^2s_3-2s_1s_2^2s_3+s_2^3s_3
\nonumber\\
&\quad
-\frac{1}{2}s_3^2+s_1s_3^2-s_1^2s_3^2
+s_2s_3^2+s_1s_2s_3^2-s_2s_3^3+\frac{1}{2}s_3^4
\nonumber\\
&\quad
+\alpha\left(s_1-s_2+s_3\right)
=0,
\end{align}
respectively, which are both satisfied by \eqref{alpha solution 2 deriv AdS vertex}.\\
\indent For the three-derivative vertex, one must solve \eqref{AdS gauge variation in first field three deriv vertex}, which initially reduces to 
\begingroup
\small
\allowdisplaybreaks[4]
\begin{align}
&-p_3\,y_1y_2^2y_3\,
z_1^{p_1}z_2^{p_2}z_3^{p_3-1}
+p_2\,y_1y_2y_3^2\,
z_1^{p_1}z_2^{p_2-1}z_3^{p_3}
\nonumber\\[1mm]
&\quad
+\frac{1}{2\ell^2}
\bigl(s_1-s_2+s_3-1\bigr)
\bigl(1+\alpha_1\bigr)
y_1y_3\,
z_1^{p_1+1}z_2^{p_2-1}z_3^{p_3}
\nonumber\\
&\quad
-\frac{1}{2\ell^2}
\bigl(s_1+s_2-s_3-1\bigr)
\bigl(s_2-s_3+\alpha_1\bigr)
y_1y_2\,
z_1^{p_1+1}z_2^{p_2}z_3^{p_3-1}
\nonumber\\
&\quad
-\frac{1}{2\ell^2}
\bigl(s_1+s_2-s_3-1\bigr)
\bigl(1+\alpha_2\bigr)
y_2^2\,
z_1^{p_1}z_2^{p_2+1}z_3^{p_3-1}
\nonumber\\
&\quad
+\frac{1}{2\ell^2}
\bigl(s_1-s_2+s_3-1\bigr)
\alpha_3\,y_3^2\,
z_1^{p_1}z_2^{p_2-1}z_3^{p_3+1}
\displaybreak[3]\nonumber\\[1mm]
&\quad
+\frac{1}{2\ell^2}
y_2y_3\,z_1^{p_1}z_2^{p_2}z_3^{p_3}
\Big[
-1+4s_1-s_1^2+3s_2+s_1s_2-6s_3-s_2s_3+s_3^2
\nonumber\\
&\hspace{3.1cm}
+\bigl(1+s_1-s_2+s_3\bigr)\alpha_2
+\bigl(-1-s_1-s_2+s_3\bigr)\alpha_3
\Big]
\displaybreak[3]\nonumber\\[1mm]
&\quad
+\frac{1}{2\ell^4}
z_1^{p_1+1}z_2^{p_2}z_3^{p_3}
\Big[
s_2+s_1s_2-s_2^2-s_3-s_1s_3-2s_2s_3+3s_3^2
\nonumber\\
&\hspace{3.1cm}
+\bigl(
-1+2s_1-s_1^2+3s_2+s_1s_2
-4s_3-s_2s_3+s_3^2
\bigr)\alpha_1
\nonumber\\
&\hspace{3.1cm}
+\bigl(
-s_2-s_1s_2-s_2^2+s_3+s_1s_3
+2s_2s_3-s_3^2
\bigr)\alpha_3
\Big].
\end{align}
\endgroup
Then we use the Schouten identity \eqref{AdS DDI 6} to trade all four-derivative terms for two and zero derivative terms, followed by \eqref{AdS DDI 1} to remove the two-derivative terms of the form $y_i^2$, which is sufficient for an independent basis. Similar to the two-derivative vertex, we get an overdetermined set of constraints. The zero-derivative term requires:
\begingroup
\small
\allowdisplaybreaks[4]
\begin{align}
&
-2+s_1+5s_1^2-7s_1^3+3s_1^4
+9s_2-11s_1s_2+8s_1^2s_2+5s_1^3s_2-s_1^4s_2
\nonumber\\
&\quad
-22s_2^2+25s_1s_2^2+s_1^2s_2^2-2s_1^3s_2^2
+16s_2^3-15s_1s_2^3
\nonumber\\
&\quad
-6s_2^4+2s_1s_2^4+s_2^5
-s_3-10s_1s_3+8s_1^2s_3-8s_1^3s_3+s_1^4s_3
\nonumber\\
&\quad
+25s_2s_3-36s_1s_2s_3+s_1^2s_2s_3
-20s_2^2s_3+6s_1s_2^2s_3
\nonumber\\
&\quad
+11s_2^3s_3-s_2^4s_3
+3s_3^2+9s_1s_3^2-6s_1^2s_3^2+2s_1^3s_3^2
\nonumber\\
&\quad
+20s_2s_3^2+7s_1s_2s_3^2-3s_2^2s_3^2-2s_2^3s_3^2
-20s_3^3
\nonumber\\
&\quad
+6s_1s_3^3-11s_2s_3^3+2s_2^2s_3^3
+9s_3^4-2s_1s_3^4+s_2s_3^4-s_3^5
\nonumber\\[1mm]
&\quad
+\Bigl(
-2+4s_1-2s_1^2+6s_2+2s_1s_2-8s_3-2s_2s_3+2s_3^2
\Bigr)\alpha_1
\nonumber\\[1mm]
&\quad
+\Bigl(
2-4s_1+2s_1^3-6s_2+6s_1s_2+2s_1^2s_2
+6s_2^2-2s_1s_2^2-2s_2^3
\nonumber\\
&\hspace{2.6cm}
+8s_3-8s_1s_3-2s_1^2s_3-10s_2s_3+2s_2^2s_3
\nonumber\\
&\hspace{2.6cm}
+4s_3^2+2s_1s_3^2+2s_2s_3^2-2s_3^3
\Bigr)\alpha_2
\nonumber\\[1mm]
&\quad
+\Bigl(
2-4s_1+4s_1^2-2s_1^3
-2s_2-2s_1s_2+2s_1^2s_2
-2s_2^2-2s_1s_2^2+2s_2^3
\nonumber\\
&\hspace{2.6cm}
+4s_3-2s_1^2s_3+10s_2s_3-2s_2^2s_3
-8s_3^2+2s_1s_3^2-2s_2s_3^2+2s_3^3
\Bigr)\alpha_3
=0,
\end{align}
\endgroup
and there are three constraints arising from the two-derivative terms:
\begingroup
\small
\allowdisplaybreaks[4]
\begin{align}
&2+s_1-4s_1^2+s_1^3
-5s_2+4s_1s_2+s_1^2s_2
+4s_2^2-s_1s_2^2-s_2^3
\nonumber\\
&\quad
-5s_3+6s_1s_3-s_1^2s_3
-8s_2s_3+s_2^2s_3
+8s_3^2-3s_1s_3^2+s_2s_3^2-s_3^3
\nonumber\\
&\quad
+2\left(1-s_1-s_2+s_3\right)\alpha_1
+2\left(-1+s_1+s_2-s_3\right)\alpha_2
\nonumber\\
&\quad
+2\left(-1+s_1-s_2+s_3\right)\alpha_3
=0 ,
\label{eq:three-derivative-alpha-constraint-1}
\\[1.5ex]
&2s_1^2+3s_1s_2-s_1^2s_2-s_1s_2^2
-5s_1s_3+s_1^2s_3+s_1s_3^2
\nonumber\\
&\quad
+2s_1\alpha_2-2s_1\alpha_3
=0 ,
\label{eq:three-derivative-alpha-constraint-2}
\\[1.5ex]
&-2-s_1+4s_1^2-s_1^3
+9s_2-14s_1s_2+s_1^2s_2
-4s_2^2+3s_1s_2^2+s_2^3
\nonumber\\
&\quad
+s_3+4s_1s_3-s_1^2s_3
+4s_2s_3-s_2^2s_3
-4s_3^2+s_1s_3^2-s_2s_3^2+s_3^3
\nonumber\\
&\quad
+2\left(-1+s_1-s_2+s_3\right)\alpha_1
+2\left(1-s_1-s_2+s_3\right)\alpha_2
\nonumber\\
&\quad
+2\left(1-s_1+s_2-s_3\right)\alpha_3
=0,
\label{eq:three-derivative-alpha-constraint-3}
\end{align}
\endgroup
arising from the $y_1y_2$, $y_2y_3$ and $y_1y_3$ coefficients respectively. Remarkably, these are all satisfied by the cyclic set of coefficients \eqref{alphai three derivative vertex 1 deriv coefficients}.
\section{Minimal Gravitational Coupling from the Fronsdal Action}
\label{Minimal Gravitational Coupling from the Fronsdal Action}
In this appendix, we explicitly verify that the AdS$_3$ correction to the flat space \(s\text{-}s\text{-}2\) cubic vertex
\begin{equation}
\mathcal{V}_0
=
-\left(
s y_1z_1+s y_2z_2+y_3z_3
\right)y_3z_3^{s-1}
\label{flat space two deriv cubic vertex}
\end{equation}
is given by
\begin{equation}
\frac{1}{\ell^2}\mathcal{V}_1
=
\frac{s(s+1)}{\ell^2}
z_1z_2z_3^{s-1}.
\label{AdS vertex correction}
\end{equation}
This vertex couples the fields
\begin{equation*}
\phi_1^{\mu_1...\mu_s},
\qquad
\phi_2^{\mu_1...\mu_s},
\qquad
h_3^{\mu\nu},
\end{equation*}
where \(\phi_1\) and \(\phi_2\) denote the two spin-\(s\) entries of the cubic vertex and \(h_3\) is the spin-two field. As mentioned in Section~\ref{Consistency check for minimal coupling to gravity}, the minimal gravitational coupling is obtained by covariantising the Fronsdal action \eqref{initial linearised AdS fronsdal lagrangian} and subsequently linearising with respect to a perturbation in the metric and its inverse \eqref{metric linearisation and inverse}.

We begin by collecting the notation and identities required for the calculation. We denote the AdS background covariant derivative by \(\bar{\nabla}\), while \(\nabla\) denotes the covariant derivative with respect to the full metric. To linear order in the spin-two perturbation \(h_{\mu\nu}\), its action on a symmetric spin-$s$ field is
\begin{align}
\nabla_{\nu}\phi_{\mu_1...\mu_s}
={}&
\bar{\nabla}_{\nu}\phi_{\mu_1...\mu_s}
-\Gamma^{\sigma}{}_{\nu\mu_1}
\phi_{\sigma\mu_2\cdots\mu_s}
-\cdots
-\Gamma^{\sigma}{}_{\nu\mu_s}
\phi_{\mu_1\cdots\mu_{s-1}\sigma}\nonumber\\
={}&\bar{\nabla}_{\nu}\phi_{\mu_1...\mu_s} - s\Gamma^{\sigma}{}_{\nu(\mu_1}\phi_{\mu_2...\mu_s)\sigma},
\label{covariant derivative on spin-s field}
\end{align}
where the linearised Christoffel symbol is
\begin{equation}
\Gamma^{\mu}{}_{\nu\rho}
=
\frac{1}{2}
\left(
\bar{\nabla}_{\nu}h_{\rho}{}^{\mu}
+\bar{\nabla}_{\rho}h_{\nu}{}^{\mu}
-\bar{\nabla}^{\mu}h_{\nu\rho}
\right).
\label{linearised christoffel symbol}
\end{equation}
We will also require the commutator of two background AdS covariant derivatives acting on a spin-\(s\) field:
\begin{align}
[\bar{\nabla}_{\rho},\bar{\nabla}_{\sigma}]\phi_{\mu_1...\mu_s}
={}&
-\left(
{R^{\nu}}_{\mu_1\rho\sigma}\phi_{\nu\mu_2\cdots\mu_s}
+\cdots+
{R^{\nu}}_{\mu_s\rho\sigma}
\phi_{\mu_1\cdots\mu_{s-1}\nu}
\right)
\nonumber\\
={}&
-s\phi_{\nu(\mu_1...\mu_{s-1}}{R^{\nu}}_{\mu_s)\rho\sigma}\nonumber\\
={}& \frac{s}{\ell^2}\left(\phi_{\rho(\mu_1...\mu_{s-1}}g_{\mu_s)\sigma} - \phi_{\sigma(\mu_1...\mu_{s-1}}g_{\mu_s)\rho}\right)
\label{commutator -> riemann tensor expression}
\end{align}
where in the last equality we used the explicit expression for the AdS Riemann tensor \eqref{eq:riemann-tensor}.\\
\indent The AdS Fronsdal Lagrangian to relevant order is
\begin{align}
\mathcal{L}_{\mathrm{Frons.}}^{\mathrm{AdS}}
\rightarrow{}&
{\phi_2}^{\mu_1...\mu_s}\Box{\phi_1}_{\mu_1...\mu_s}
-\frac{s(s-3)}{\ell^2}
{\phi_2}^{\mu_1...\mu_s}{\phi_1}_{\mu_1...\mu_s}
\nonumber\\
&\quad
+\mathcal{O}\!\left((\nabla\cdot\phi)^2\right)
+\mathcal{O}\!\left((\operatorname{Tr}\phi)^2\right)
+\mathcal{O}\!\left(\operatorname{Tr}(\phi)\,(\nabla\cdot\phi)\right),
\label{initial linearised AdS fronsdal lagrangian no sqrt(-g)}
\end{align}
where the metric determinant factor $\sqrt{-g}$ in \eqref{initial linearised AdS fronsdal lagrangian} has been set to one at the linearised TT level because
\begin{equation}
\sqrt{-g} = \sqrt{-det\left(\eta+h\right)} = 1 + \frac{1}{2}{h^{\mu}}_{\mu} + \mathcal{O}(h^2) \overset{\mathrm{TT}}{\approx} 1.
\label{linearised metric determinant}
\end{equation}
We first show that the cubic terms generated by expanding the metric contractions between the spin-$s$ fields do not contribute. After inserting a metric for one of the contracted field indices in \eqref{initial linearised AdS fronsdal lagrangian no sqrt(-g)}, it becomes
\begin{align}
\mathcal{L}_{\mathrm{Frons.}}^{\mathrm{AdS}}
\rightarrow{}&
\left(\eta^{\nu\sigma}-h_3^{\nu\sigma}\right)
{\phi_2}_{\nu}{}^{\mu_2\cdots\mu_s}
\Box{\phi_1}_{\sigma\mu_2\cdots\mu_s}
\nonumber\\
&\quad
-\frac{s(s-3)}{\ell^2}
\left(\eta^{\nu\sigma}-h_3^{\nu\sigma}\right)
{\phi_2}_{\nu}{}^{\mu_2\cdots\mu_s}
{\phi_1}_{\sigma\mu_2\cdots\mu_s}.
\label{metric contraction expansion in AdS lagrangian}
\end{align}
where in the second order trace/divergence terms in \eqref{initial linearised AdS fronsdal lagrangian no sqrt(-g)}, at least one divergence or trace will persist, and so are dropped in the TT sector. Then, one can simply eliminate the $\mathcal{O}(\phi\phi h)$ terms by field redefinition $\phi_2^{\mu_1...\mu_s} \rightarrow \phi_2^{\mu_1...\mu_s} + {h_3}^{\nu(\mu_1}{{\phi_2}_{\nu}}^{\mu_2...\mu_s)}$:
\begin{multline}
\mathcal{L}_{\mathrm{Frons.}}^{\mathrm{AdS}} \rightarrow \eta^{\nu \sigma}{{\phi_2}_{\nu}}^{\mu_2...\mu_s}\Box{\phi_1}_{\sigma\mu_2...\mu_s} - \frac{s(s-3)}{\ell^2}{\eta^{\nu \sigma}}{{\phi_2}_{\nu}}^{\mu_2...\mu_s}{\phi_1}_{\sigma\mu_2...\mu_s} \\ -{h_3}^{\nu \sigma}\left(\cancelto{\scriptstyle 0, \text{ by field redefinition}} {{{\phi_2}_{\nu}}^{\mu_2...\mu_s}\Box{\phi_1}_{\sigma\mu_2...\mu_s}  - \frac{s(s-3)}{\ell^2}{{\phi_2}_{\nu}}^{\mu_2...\mu_s}{\phi_1}_{\sigma\mu_2...\mu_s}} \right) + \mathcal{O}(\phi\phi h^2)\\ 
\end{multline}
\begin{equation}
  \implies \mathcal{L}_{\mathrm{Frons.}}^{\mathrm{AdS}} \overset{\mathcal{O}(h^1)}{\approx} {\phi_2}^{\mu_1...\mu_s}\Box{\phi_1}_{\mu_1...\mu_s},
  \label{fronsdal AdS action box term only relevant for hphiphi vertex}
\end{equation}
where the final line also ignores the AdS$_3$ mass term, as it is $\mathcal{O}(h^0)$ and therefore not relevant to our analysis. To expand \eqref{fronsdal AdS action box term only relevant for hphiphi vertex}, we pull out the metric contracting the box operator
\begin{align}
\mathcal{L}_{\mathrm{Frons.}}^{\mathrm{AdS}} \rightarrow{}&{\phi_2}^{\mu_1...\mu_s}{\eta^{\alpha \beta}}\nabla_{\alpha} \nabla_{\beta}\,{\phi_1}_{\mu_1...\mu_s} - {\phi_2}^{\mu_1...\mu_s}{h_3}^{\alpha \beta}\nabla_{\alpha} \nabla_{\beta}\,{\phi_1}_{\mu_1...\mu_s}.
\label{insert metric for box contraction}
\end{align}
The second term in \eqref{insert metric for box contraction} is readily simplified by dropping $\mathcal{O}(h\Gamma)$ terms in the expansion of the covariant derivative, and then noticing that it already has the desired form:
\begin{align}
{\phi_2}^{\mu_1...\mu_s}{h_3}^{\alpha \beta}\nabla_{\alpha} \nabla_{\beta}\,{\phi_1}_{\mu_1...\mu_s}
&= {\phi_2}^{\mu_1...\mu_s}{h_3}^{\alpha \beta}\bar{\nabla}_{\alpha} \bar{\nabla}_{\beta}\,{\phi_1}_{\mu_1...\mu_s} + \mathcal{O}(\phi\phi h^2)\\
&= (\partial_3\cdot \nabla_1)^2(\partial_1\cdot \partial_2)^s\phi_1(a)\phi_2(a)h_3(a)
      + \mathcal{O}(\phi\phi h^2) \\
&= y_3^2{z_3}^s\phi_1(a)\phi_2(a)h_3(a) + \mathcal{O}(\phi\phi h^2).
\label{y3^2z3^s term from expansion of box in AdS frondsdal lagrangian}
\end{align}
This corresponds to the third term inside the brackets in \eqref{flat space two deriv cubic vertex}. Considering now the first term in \eqref{insert metric for box contraction}, we can expand the covariant derivatives as follows,
\begin{align}
{\phi_2}^{\mu_1...\mu_s}{\eta^{\alpha \beta}}\nabla_{\alpha} \nabla_{\beta}\,{\phi_1}_{\mu_1...\mu_s}
  &= \eta^{\alpha\beta}{\phi_2}^{\mu_1...\mu_s}
  \Bigl( \bar\nabla_{\alpha}\nabla_{\beta}{\phi_1}_{\mu_1...\mu_s} \nonumber\\[1mm]
  &\quad - \Gamma^{\xi}_{\alpha\beta}\nabla_{\xi}{\phi_1}_{\mu_1...\mu_s}
  - s\,\Gamma^{\xi}_{\alpha\mu_1}\nabla_{\beta}{\phi_1}_{\xi\mu_2...\mu_s}\Bigr)\\[1mm]
  &= \eta^{\alpha \beta}{\phi_2}^{\mu_1...\mu_s}
  \Bigl[ \bar\nabla_{\alpha}\bar\nabla_{\beta}{\phi_1}_{\mu_1...\mu_s} \nonumber\\[1mm]
  &\quad - s\,\bar\nabla_{\alpha}\left(\Gamma^{\xi}_{\beta\mu_1}\,{\phi_1}_{\xi\mu_2...\mu_s}\right)
  - \Gamma^{\xi}_{\alpha\beta}\bar\nabla_{\xi}{\phi_1}_{\mu_1...\mu_s} \nonumber\\[1mm]
  &\quad - s\,\Gamma^{\xi}_{\alpha\mu_1}\bar\nabla_{\beta}{\phi_1}_{\xi\mu_2...\mu_s}\Bigr]
  + \mathcal{O}(\Gamma^2 \sim h^2)
\end{align}
\begin{equation*}
\mathbin{\overset{\mathcal{O}(h^1)}{\approx} -\eta^{\alpha\beta}{\phi_2}^{\mu_1...\mu_s}
  \Bigl[2s\,\Gamma^{\xi}_{\alpha\mu_1}\bar\nabla_{\beta}{\phi_1}_{\xi\mu_2...\mu_s}
  + \Gamma^{\xi}_{\alpha\beta}\bar\nabla_{\xi}{\phi_1}_{\mu_1...\mu_s}
  + s\,\Bigl(\bar\nabla_{\alpha}\Gamma^{\xi}_{\beta\mu_1}\Bigr){\phi_1}_{\xi\mu_2...\mu_s}\Bigr]},
\end{equation*}
where we dropped the leading $\mathcal{O}(h^0)$ term. Note, we used the symmetry of the field indices to avoid writing out the explicit symmetrisation $(...)$ in our use of \eqref{covariant derivative on spin-s field}, as well as to combine the $\sim\Gamma\nabla\phi$ terms in the last line. Now we substitute the linearised Christoffel symbols \eqref{linearised christoffel symbol} and contract the background metrics,
\begin{multline}
\mathcal{L}_{\mathrm{Frons.}}^{\mathrm{AdS}} \xrightarrow{\text{no } \sim\,y_3^2 \text{ \ref{y3^2z3^s term from expansion of box in AdS frondsdal lagrangian} term}} -\frac{1}{2}{\phi_2}^{\mu_1...\mu_s}
  \Bigl[2s\,(\overset{\textcircled{1}}{\overbrace{{\bar{\nabla}}_{\alpha} {{h_3}_{\mu_1}}^{\xi}}} + \overset{\textcircled{2}}{\overbrace{{\bar{\nabla}}_{\mu_1} {{h_3}_{\alpha}}^{\xi}}} - \overset{\textcircled{3}}{\overbrace{{\bar{\nabla}}^{\xi}{h_3}_{\alpha\mu_1}}})\bar\nabla^{\alpha}{\phi_1}_{\xi\mu_2...\mu_s}
  \\+\,\cancelto{\scriptstyle \mathcal{O}(h^2) \text{ by TT }}{\left(2{\bar{\nabla}}^{\alpha} {{h_3}_{\alpha}}^{\xi} - {\bar{\nabla}}^{\xi}{{h_3}_{\alpha}}^{\alpha}\right)}\bar\nabla_{\xi}{\phi_1}_{\mu_1...\mu_s}
  + s\,(\bar\nabla^{\alpha}(\underset{\textcircled{4}}{\underbrace{{\bar{\nabla}}_{\alpha} {{h_3}_{\mu_1}}^{\xi}}} + \underset{\textcircled{5}}{\underbrace{{{\bar{\nabla}}_{\mu_1} {{h_3}_{\alpha}}^{\xi}}}} - \underset{\textcircled{6}}{\underbrace{{{\bar{\nabla}}^{\xi}{h_3}_{\alpha\mu_1}}}})){\phi_1}_{\xi\mu_2...\mu_s}\Bigr],
\end{multline}
where it should be noted that for these terms, $\eta$ contractions are equivalent to full contractions up to $\mathcal{O}(h^2)$. We then manually bring the expression into the minimal operator basis, treating each of the numbered terms above separately. Starting with:
\begin{multline}
\textcircled{1} = -s{\phi_2}^{\mu_1...\mu_s}\bar\nabla_{\alpha}{{h_3}_{\mu_1}}^{\xi}\bar\nabla^{\alpha}{\phi_1}_{\xi\mu_2...\mu_s} \overset{\text{IBP}}{=} \frac{1}{2}s{\phi_2}^{\sigma\mu_2...\mu_s}{{h_3}_{\sigma}}^{\xi}\bar\Box{\phi_1}_{\xi\mu_2...\mu_s}
\\ + \frac{1}{2}s{\phi_2}^{\sigma\mu_2...\mu_s}\bar\Box{{h_3}_{\sigma}}^{\xi}{\phi_1}_{\xi\mu_2...\mu_s} - \frac{1}{2}s\bar\Box{\phi_2}^{\sigma\mu_2...\mu_s}{{h_3}_{\sigma}}^{\xi}{\phi_1}_{\xi\mu_2...\mu_s},
\end{multline}
the on-shell condition,
\begin{equation}
\bar\Box\phi_{\mu(s)} = \frac{s(s-3)}{\ell^2}\phi_{\mu(s)},
\end{equation}
causes two terms to cancel, leaving
\begin{equation}
\textcircled{1} = -\frac{s}{\ell^2}{\phi_2}^{\sigma\mu_2...\mu_s}{\phi_1}_{\xi\mu_2...\mu_s} {{h_3}_{\sigma}}^{\xi} = -\frac{s}{\ell^2}z_1z_2z_3^{s-1}\phi_1(a)\phi_2(a)h_3(a).
\end{equation}
The next term, \textcircled{2}, is already in the desired form,
\begin{equation}
   \textcircled{2} = -s{{\phi_2}^{\sigma}}^{\mu_2...\mu_s}{\bar{\nabla}}_{\sigma} {{h_3}_{\alpha}}^{\xi} \bar\nabla^{\alpha}{\phi_1}_{\xi\mu_2...\mu_s} = -s y_2y_3 z_2{z_3}^{s-1}\phi_1(a)\phi_2(a)h_3(a),
\end{equation}
which is the second term inside the brackets in \eqref{flat space two deriv cubic vertex}, while \textcircled{3} requires commutators of the background derivatives,
\begin{align}
    \textcircled{3} ={}&s{\phi_2}^{\sigma\mu_2...\mu_s}{\bar{\nabla}}^{\xi}{h_3}_{\alpha\sigma}\bar\nabla^{\alpha}{\phi_1}_{\xi\mu_2...\mu_s}\\
    \overset{\text{IBP}}{=}& -s{\bar{\nabla}}^{\xi}{\phi_2}^{\sigma\mu_2...\mu_s}{h_3}_{\alpha\sigma}\bar\nabla^{\alpha}{\phi_1}_{\xi\mu_2...\mu_s} - s{\phi_2}^{\sigma\mu_2...\mu_s}{h_3}_{\alpha\sigma}{\bar{\nabla}}^{\xi}\bar\nabla^{\alpha}{\phi_1}_{\xi\mu_2...\mu_s}\\
    ={}& -sy_1y_3z_1{z_3}^{s-1}\phi_1(a)\phi_2(a)h_3(a) - s{{\phi_2}^{\sigma}}^{\mu_2...\mu_s}{h_3}_{\alpha\sigma}[{\bar{\nabla}}^{\xi},\bar\nabla^{\alpha}]{\phi_1}_{\xi\mu_2...\mu_s},
\end{align}
giving the final term in \eqref{flat space two deriv cubic vertex}. The zero derivative contribution becomes
\begin{align}
& s{{\phi_2}^{\sigma}}^{\mu_2...\mu_s}{h_3}_{\alpha\sigma} \left[{{R^{\rho}}_{\xi}}^{\xi\alpha}{\phi_1}_{\rho\mu_2...\mu_s} + (s-1) {{R^{\rho}}_{\mu_2}}^{\xi\alpha}{\phi_1}_{\xi\rho\mu_3...\mu_s}\right] \\
     &= s{\phi_2}^{\sigma\mu_2...\mu_s}{h_3}_{\alpha\sigma} \left[\frac{2}{\ell^2}g^{\rho\alpha}{\phi_1}_{\rho\mu_2...\mu_s} - \frac{1}{\ell^2}(s-1) ( \cancelto{\scriptstyle \mathcal{O}(h^2) \text{ by TT }}{g^{\rho\xi}\delta_{\mu_2}^{\alpha}} - g^{\rho\alpha}\delta_{\mu_2}^{\xi}){\phi_1}_{\xi\rho\mu_3...\mu_s}\right] \\
     &\overset{\mathcal{O}(h^1)}{\approx} \frac{s(s+1)}{\ell^2}{\phi_2}^{\sigma\mu_2...\mu_s}{h_3}_{\alpha\sigma}{{\phi_1}^{\alpha}}_{\mu_2...\mu_s} = \frac{s(s+1)}{\ell^2}z_1z_2{z_3}^{s-1}\phi_1(a)\phi_2(a)h_3(a)
\label{surviving contribution to zero derivative coefficient}
\end{align}
The remaining terms can be calculated using the same principles:
\begin{align}
    \textcircled{4} &= -\frac{s}{2}{\phi_1}_{\xi\mu_2...\mu_s}{\phi_2}^{\sigma\mu_2...\mu_s}\bar\nabla^{\alpha}{\bar{\nabla}}_{\alpha}{{h_3}_{\sigma}}^{\xi} = \frac{s}{\ell^2}z_1z_2z_3^{s-1}\phi_1(a)\phi_2(a)h_3(a)
\end{align}
\begin{align}
    \textcircled{5} &= -\frac{s}{2}{\phi_1}_{\xi\mu_2...\mu_s}{\phi_2}^{\sigma\mu_2...\mu_s}\bar\nabla^{\alpha}{\bar{\nabla}}_{\sigma}{{h_3}_{\alpha}}^{\xi} \\ 
    &= -\frac{s}{2}{\phi_1}_{\xi\mu_2...\mu_s}{{\phi_2}^{\sigma}}^{\mu_2...\mu_s}[\bar\nabla_{\alpha},{\bar{\nabla}}_{\sigma}]{{h_3}^{\alpha}}^{\xi}\\ 
    &= -\frac{s}{2}{\phi_1}_{\xi\mu_2...\mu_s}{\phi_2}^{\sigma\mu_2...\mu_s}\left({R^{\alpha}}_{\nu\alpha\sigma}{h_3}^{\nu\xi} + {R^{\xi}}_{\nu\alpha\sigma}{h_3}^{\alpha\nu}\right)\\
    &= \frac{s}{2}{\phi_1}_{\xi\mu_2...\mu_s}{{\phi_2}^{\sigma}}^{\mu_2...\mu_s}\left[\frac{2}{\ell^2}g_{\sigma\nu}{h_3}^{\nu\xi} - \frac{1}{\ell^2}(\cancelto{\scriptstyle \mathcal{O}(h^2) \text{ by TT }}{\delta^{\xi}_{\sigma} g_{\nu\alpha}} - \delta^{\xi}_{\alpha} g_{\nu\sigma}){h_3}^{\nu\alpha}\right]\\
    &= \frac{3s}{2\ell^2}{\phi_1}_{\xi\mu_2...\mu_s}{\phi_2}^{\sigma\mu_2...\mu_s}{{h_3}_{\sigma}}^{\xi} = \frac{3s}{2\ell^2}z_1z_2z_3^{s-1}\phi_1(a)\phi_2(a)h_3(a)
\end{align}
\begin{align}
    \textcircled{6} &= \frac{s}{2}{\phi_1}_{\xi\mu_2...\mu_s}{{\phi_2}^{\sigma}}^{\mu_2...\mu_s}\bar\nabla^{\alpha}{\bar{\nabla}}^{\xi}{{h_3}_{\alpha}}_{\sigma}\\&= -\frac{3s}{2\ell^2}z_1z_2z_3^{s-1}\phi_1(a)\phi_2(a)h_3(a).
\end{align}
Result \textcircled{6} follows due to equivalence with \textcircled{5} up to a sign under the exchange $1\leftrightarrow2$. We see, then, that \textcircled{1} cancels with \textcircled{4}, and \textcircled{5} cancels with \textcircled{6}, leaving \eqref{surviving contribution to zero derivative coefficient} as the only zero-derivative contribution to the cubic vertex, in agreement with \eqref{AdS vertex correction}.
\bibliographystyle{JHEP}
\bibliography{References}

@article{Kessel:2018ugi,
    author = "Kessel, Pan and Mkrtchyan, Karapet",
    title = "{Cubic interactions of massless bosonic fields in three dimensions II: Parity-odd and Chern-Simons vertices}",
    eprint = "1803.02737",
    archivePrefix = "arXiv",
    primaryClass = "hep-th",
    doi = "10.1103/PhysRevD.97.106021",
    journal = "Phys. Rev. D",
    volume = "97",
    number = "10",
    pages = "106021",
    year = "2018"
}

@article{Fierz:1939ix,
    author = "Fierz, M. and Pauli, W.",
    title = "{On relativistic wave equations for particles of arbitrary spin in an electromagnetic field}",
    doi = "10.1098/rspa.1939.0140",
    journal = "Proc. Roy. Soc. Lond. A",
    volume = "173",
    pages = "211--232",
    year = "1939"
}

@article{Fredenhagen:2024lps,
    author = "Fredenhagen, Stefan and Lausch, Filipp and Mkrtchyan, Karapet",
    title = "{Interactions of massless fermionic fields in three dimensions}",
    eprint = "2404.00497",
    archivePrefix = "arXiv",
    primaryClass = "hep-th",
    doi = "10.1103/PhysRevD.110.L081702",
    journal = "Phys. Rev. D",
    volume = "110",
    number = "8",
    pages = "L081702",
    year = "2024"
}

@article{Sharapov:2024euk,
    author = "Sharapov, Alexey and Skvortsov, Evgeny and Sukhanov, Arseny",
    title = "{Matter-coupled higher spin gravities in 3d: no- and yes-go results}",
    eprint = "2409.12830",
    archivePrefix = "arXiv",
    primaryClass = "hep-th",
    doi = "10.1007/JHEP04(2025)155",
    journal = "JHEP",
    volume = "04",
    pages = "155",
    year = "2025"
}

@article{Gubser:1998bc,
    author = "Gubser, S. S. and Klebanov, Igor R. and Polyakov, Alexander M.",
    title = "{Gauge theory correlators from noncritical string theory}",
    eprint = "hep-th/9802109",
    archivePrefix = "arXiv",
    reportNumber = "PUPT-1767",
    doi = "10.1016/S0370-2693(98)00377-3",
    journal = "Phys. Lett. B",
    volume = "428",
    pages = "105--114",
    year = "1998"
}

@article{Witten:1998qj,
    author = "Witten, Edward",
    title = "{Anti de Sitter space and holography}",
    eprint = "hep-th/9802150",
    archivePrefix = "arXiv",
    reportNumber = "IASSNS-HEP-98-15",
    doi = "10.4310/ATMP.1998.v2.n2.a2",
    journal = "Adv. Theor. Math. Phys.",
    volume = "2",
    pages = "253--291",
    year = "1998"
}

@article{Sezgin:2002rt,
    author = "Sezgin, E. and Sundell, P.",
    title = "{Massless higher spins and holography}",
    eprint = "hep-th/0205131",
    archivePrefix = "arXiv",
    reportNumber = "CTP-TAMU-08-02, UU-01-10",
    doi = "10.1016/S0550-3213(02)00739-3",
    journal = "Nucl. Phys. B",
    volume = "644",
    pages = "303--370",
    year = "2002",
    note = "[Erratum: Nucl.Phys.B 660, 403--403 (2003)]"
}

@article{Giombi:2012ms,
    author = "Giombi, Simone and Yin, Xi",
    title = "{The Higher Spin/Vector Model Duality}",
    eprint = "1208.4036",
    archivePrefix = "arXiv",
    primaryClass = "hep-th",
    doi = "10.1088/1751-8113/46/21/214003",
    journal = "J. Phys. A",
    volume = "46",
    pages = "214003",
    year = "2013"
}

@inproceedings{Bekaert:2004qos,
    author = "Bekaert, X. and Cnockaert, S. and Iazeolla, Carlo and Vasiliev, M. A.",
    title = "{Nonlinear higher spin theories in various dimensions}",
    booktitle = "{1st Solvay Workshop on Higher Spin Gauge Theories}",
    eprint = "hep-th/0503128",
    archivePrefix = "arXiv",
    reportNumber = "IHES-P-04-47, ULB-TH-04-26, ROM2F-04-29, FIAN-TD-17-04, IHES/P/04/47, ULB-TH/04-26, ROM2F-04/29, FIAN/TD/17/04",
    pages = "132--197",
    year = "2004"
}

@article{Boulanger:2008tg,
    author = "Boulanger, Nicolas and Leclercq, Serge and Sundell, Per",
    title = "{On The Uniqueness of Minimal Coupling in Higher-Spin Gauge Theory}",
    eprint = "0805.2764",
    archivePrefix = "arXiv",
    primaryClass = "hep-th",
    doi = "10.1088/1126-6708/2008/08/056",
    journal = "JHEP",
    volume = "08",
    pages = "056",
    year = "2008"
}

@article{Berends:1984rq,
    author = "Berends, Frits A. and Burgers, G. J. H. and van Dam, H.",
    title = "{On the Theoretical Problems in Constructing Interactions Involving Higher Spin Massless Particles}",
    reportNumber = "IFP-234-UNC",
    doi = "10.1016/0550-3213(85)90074-4",
    journal = "Nucl. Phys. B",
    volume = "260",
    pages = "295--322",
    year = "1985"
}

@article{Sundborg:2000wp,
    author = "Sundborg, Bo",
    editor = "Sorokin, Dmitri P.",
    title = "{Stringy gravity, interacting tensionless strings and massless higher spins}",
    eprint = "hep-th/0103247",
    archivePrefix = "arXiv",
    doi = "10.1016/S0920-5632(01)01545-6",
    journal = "Nucl. Phys. B Proc. Suppl.",
    volume = "102",
    pages = "113--119",
    year = "2001"
}

@article{Maldacena:1997re,
    author = "Maldacena, Juan Martin",
    title = "{The Large $N$ limit of superconformal field theories and supergravity}",
    eprint = "hep-th/9711200",
    archivePrefix = "arXiv",
    reportNumber = "HUTP-97-A097, HUTP-98-A097",
    doi = "10.4310/ATMP.1998.v2.n2.a1",
    journal = "Adv. Theor. Math. Phys.",
    volume = "2",
    pages = "231--252",
    year = "1998"
}

@article{Bonelli:2003kh,
    author = "Bonelli, Giulio",
    title = "{On the tensionless limit of bosonic strings, infinite symmetries and higher spins}",
    eprint = "hep-th/0305155",
    archivePrefix = "arXiv",
    reportNumber = "ULB-TH-03-19",
    doi = "10.1016/j.nuclphysb.2003.07.002",
    journal = "Nucl. Phys. B",
    volume = "669",
    pages = "159--172",
    year = "2003"
}

@article{Witten:1988hc,
    author = "Witten, Edward",
    title = "{(2+1)-Dimensional Gravity as an Exactly Soluble System}",
    reportNumber = "IASSNS-HEP-88-32",
    doi = "10.1016/0550-3213(88)90143-5",
    journal = "Nucl. Phys. B",
    volume = "311",
    pages = "46",
    year = "1988"
}

@article{Fronsdal:1978vb,
    author = "Fronsdal, Christian",
    title = "{Singletons and Massless, Integral Spin Fields on de Sitter Space (Elementary Particles in a Curved Space 7).}",
    reportNumber = "UCLA/78/TEP/23",
    doi = "10.1103/PhysRevD.20.848",
    journal = "Phys. Rev. D",
    volume = "20",
    pages = "848--856",
    year = "1979"
}

@article{Joung:2011ww,
    author = "Joung, Euihun and Taronna, Massimo",
    title = "{Cubic interactions of massless higher spins in (A)dS: metric-like approach}",
    eprint = "1110.5918",
    archivePrefix = "arXiv",
    primaryClass = "hep-th",
    doi = "10.1016/j.nuclphysb.2012.03.013",
    journal = "Nucl. Phys. B",
    volume = "861",
    pages = "145--174",
    year = "2012"
}

@article{Mkrtchyan:2017ixk,
    author = "Mkrtchyan, Karapet",
    title = "{Cubic interactions of massless bosonic fields in three dimensions}",
    eprint = "1712.10003",
    archivePrefix = "arXiv",
    primaryClass = "hep-th",
    doi = "10.1103/PhysRevLett.120.221601",
    journal = "Phys. Rev. Lett.",
    volume = "120",
    number = "22",
    pages = "221601",
    year = "2018"
}

@article{Gaberdiel:2010pz,
    author = "Gaberdiel, Matthias R. and Gopakumar, Rajesh",
    title = "{An AdS$_{3}$ Dual for Minimal Model CFTs}",
    eprint = "1011.2986",
    archivePrefix = "arXiv",
    primaryClass = "hep-th",
    doi = "10.1103/PhysRevD.83.066007",
    journal = "Phys. Rev. D",
    volume = "83",
    pages = "066007",
    year = "2011"
}

@article{Fronsdal:1978rb,
    author = "Fronsdal, Christian",
    title = "{Massless Fields with Integer Spin}",
    reportNumber = "UCLA/78/TEP/5",
    doi = "10.1103/PhysRevD.18.3624",
    journal = "Phys. Rev. D",
    volume = "18",
    pages = "3624",
    year = "1978"
}

@article{Kessel:2017mxa,
    author = "Kessel, Pan",
    title = "{The Very Basics of Higher-Spin Theory}",
    eprint = "1702.03694",
    archivePrefix = "arXiv",
    primaryClass = "hep-th",
    doi = "10.22323/1.296.0001",
    journal = "PoS",
    volume = "Modave2016",
    pages = "001",
    year = "2017"
}

@article{Sagnotti:2003qa,
    author = "Sagnotti, A. and Tsulaia, M.",
    title = "{On higher spins and the tensionless limit of string theory}",
    eprint = "hep-th/0311257",
    archivePrefix = "arXiv",
    reportNumber = "ROM2F-03-31, DFPD-03-TH-43",
    doi = "10.1016/j.nuclphysb.2004.01.024",
    journal = "Nucl. Phys. B",
    volume = "682",
    pages = "83--116",
    year = "2004"
}

@article{Weinberg:1964ew,
    author = "Weinberg, Steven",
    title = "{Photons and Gravitons in  $S$-Matrix Theory: Derivation of Charge Conservation and Equality of Gravitational and Inertial Mass}",
    doi = "10.1103/PhysRev.135.B1049",
    journal = "Phys. Rev.",
    volume = "135",
    pages = "B1049--B1056",
    year = "1964"
}

@article{Aragone:1979hx,
    author = "Aragone, C. and Deser, Stanley",
    title = "{Consistency Problems of Hypergravity}",
    reportNumber = "Print-79-0377 (BRANDEIS)",
    doi = "10.1016/0370-2693(79)90808-6",
    journal = "Phys. Lett. B",
    volume = "86",
    pages = "161--163",
    year = "1979"
}

@article{Ponomarev:2022vjb,
    author = "Ponomarev, Dmitry",
    title = "{Basic Introduction to Higher-Spin Theories}",
    eprint = "2206.15385",
    archivePrefix = "arXiv",
    primaryClass = "hep-th",
    doi = "10.1007/s10773-023-05399-5",
    journal = "Int. J. Theor. Phys.",
    volume = "62",
    number = "7",
    pages = "146",
    year = "2023"
}

@mastersthesis{Neckam:2023,
    author = "Neckam, Philipp",
    title = "Spin-3-Gravity in Three-Dimensional Flat Space",
    school = "University of Vienna",
    year = "2023",
    doi= "10.25365/thesis.74353"
}

@article{Campoleoni:2024ced,
    author = "Campoleoni, Andrea and Fredenhagen, Stefan",
    title = "{Higher-Spin Gauge Theories in Three Spacetime Dimensions}",
    eprint = "2403.16567",
    archivePrefix = "arXiv",
    primaryClass = "hep-th",
    doi = "10.1007/978-3-031-59656-8_2",
    journal = "Lect. Notes Phys.",
    volume = "1028",
    pages = "121--267",
    year = "2024"
}

@article{Klebanov:2002ja,
    author = "Klebanov, I. R. and Polyakov, A. M.",
    title = "{AdS dual of the critical O(N) vector model}",
    eprint = "hep-th/0210114",
    archivePrefix = "arXiv",
    reportNumber = "PUPT-2053",
    doi = "10.1016/S0370-2693(02)02980-5",
    journal = "Phys. Lett. B",
    volume = "550",
    pages = "213--219",
    year = "2002"
}

@article{Gaberdiel:2012uj,
    author = "Gaberdiel, Matthias R. and Gopakumar, Rajesh",
    title = "{Minimal Model Holography}",
    eprint = "1207.6697",
    archivePrefix = "arXiv",
    primaryClass = "hep-th",
    doi = "10.1088/1751-8113/46/21/214002",
    journal = "J. Phys. A",
    volume = "46",
    pages = "214002",
    year = "2013"
}

@article{Belavin:1984vu,
    author = "Belavin, A. A. and Polyakov, Alexander M. and Zamolodchikov, A. B.",
    editor = "Khalatnikov, I. M. and Mineev, V. P.",
    title = "{Infinite Conformal Symmetry in Two-Dimensional Quantum Field Theory}",
    reportNumber = "CERN-TH-3827",
    doi = "10.1016/0550-3213(84)90052-X",
    journal = "Nucl. Phys. B",
    volume = "241",
    pages = "333--380",
    year = "1984"
}

@article{Fang:1978wz,
    author = "Fang, J. and Fronsdal, C.",
    title = "{Massless Fields with Half Integral Spin}",
    reportNumber = "UCLA/78/TEP/14",
    doi = "10.1103/PhysRevD.18.3630",
    journal = "Phys. Rev. D",
    volume = "18",
    pages = "3630",
    year = "1978"
}

@article{Aragone:1983sz,
    author = "Aragone, C. and Deser, Stanley",
    title = "{Hypersymmetry in $D=3$ of Coupled Gravity Massless Spin 5/2 System}",
    reportNumber = "BRX-TH-147, CIDA/TH-41",
    doi = "10.1088/0264-9381/1/2/001",
    journal = "Class. Quant. Grav.",
    volume = "1",
    pages = "L9",
    year = "1984"
}

@article{Bekaert:2010hw,
    author = "Bekaert, Xavier and Boulanger, Nicolas and Sundell, Per",
    title = "{How higher-spin gravity surpasses the spin two barrier: no-go theorems versus yes-go examples}",
    eprint = "1007.0435",
    archivePrefix = "arXiv",
    primaryClass = "hep-th",
    doi = "10.1103/RevModPhys.84.987",
    journal = "Rev. Mod. Phys.",
    volume = "84",
    pages = "987--1009",
    year = "2012"
}

@article{Vasiliev:1990en,
    author = "Vasiliev, Mikhail A.",
    title = "{Consistent equations for interacting gauge fields of all spins in (3+1)-dimensions}",
    reportNumber = "LEBEDEV-90-29",
    doi = "10.1016/0370-2693(90)91400-6",
    journal = "Phys. Lett. B",
    volume = "243",
    pages = "378--382",
    year = "1990"
}

@article{Manvelyan:2010jr,
    author = "Manvelyan, Ruben and Mkrtchyan, Karapet and Ruhl, Werner",
    title = "{General trilinear interaction for arbitrary even higher spin gauge fields}",
    eprint = "1003.2877",
    archivePrefix = "arXiv",
    primaryClass = "hep-th",
    doi = "10.1016/j.nuclphysb.2010.04.019",
    journal = "Nucl. Phys. B",
    volume = "836",
    pages = "204--221",
    year = "2010"
}

@article{Manvelyan:2010je,
    author = "Manvelyan, Ruben and Mkrtchyan, Karapet and Ruehl, Werner",
    title = "{A Generating function for the cubic interactions of higher spin fields}",
    eprint = "1009.1054",
    archivePrefix = "arXiv",
    primaryClass = "hep-th",
    doi = "10.1016/j.physletb.2010.12.049",
    journal = "Phys. Lett. B",
    volume = "696",
    pages = "410--415",
    year = "2011"
}

@article{Francia:2016weg,
    author = "Francia, Dario and Monaco, Gabriele Lo and Mkrtchyan, Karapet",
    title = "{Cubic interactions of Maxwell-like higher spins}",
    eprint = "1611.00292",
    archivePrefix = "arXiv",
    primaryClass = "hep-th",
    doi = "10.1007/JHEP04(2017)068",
    journal = "JHEP",
    volume = "04",
    pages = "068",
    year = "2017"
}

@article{Conde_2016,
   title={Spinor-helicity three-point amplitudes from local cubic interactions},
   volume={2016},
   ISSN={1029-8479},
   url={http://dx.doi.org/10.1007/JHEP08(2016)040},
   DOI={10.1007/jhep08(2016)040},
   number={8},
   journal={Journal of High Energy Physics},
   publisher={Springer Science and Business Media LLC},
   author={Conde, Eduardo and Joung, Euihun and Mkrtchyan, Karapet},
   year={2016},
   month=aug }

@article{Joung:2012fv,
    author = "Joung, Euihun and Lopez, Luca and Taronna, Massimo",
    title = "{Solving the Noether procedure for cubic interactions of higher spins in (A)dS}",
    eprint = "1207.5520",
    archivePrefix = "arXiv",
    primaryClass = "hep-th",
    doi = "10.1088/1751-8113/46/21/214020",
    journal = "J. Phys. A",
    volume = "46",
    pages = "214020",
    year = "2013"
}

@article{Achucarro:1986uwr,
    author = "Achucarro, A. and Townsend, P. K.",
    editor = "Salam, A. and Sezgin, E.",
    title = "{A Chern-Simons Action for Three-Dimensional anti-De Sitter Supergravity Theories}",
    reportNumber = "Print-87-0078 (CAMBRIDGE)",
    doi = "10.1016/0370-2693(86)90140-1",
    journal = "Phys. Lett. B",
    volume = "180",
    pages = "89",
    year = "1986"
}

@article{Blencowe:1988gj,
    author = "Blencowe, M. P.",
    title = "{A Consistent Interacting Massless Higher Spin Field Theory in $D$ = (2+1)}",
    reportNumber = "IMPERIAL/TP-87-88/30",
    doi = "10.1088/0264-9381/6/4/005",
    journal = "Class. Quant. Grav.",
    volume = "6",
    pages = "443",
    year = "1989"
}

@article{Vasiliev:1992gr,
    author = "Vasiliev, Mikhail A.",
    title = "{Unfolded representation for relativistic equations in (2+1) anti-De Sitter space}",
    reportNumber = "GOTEBORG-92-11",
    doi = "10.1088/0264-9381/11/3/015",
    journal = "Class. Quant. Grav.",
    volume = "11",
    pages = "649--664",
    year = "1994"
}

@inproceedings{Prokushkin:1998vn,
    author = "Prokushkin, Sergey and Vasiliev, Mikhail A.",
    title = "{3-d higher spin gauge theories with matter}",
    booktitle = "{2nd International Seminar on Supersymmetries and Quantum Symmetries}: {Dedicated to the Memory of Victor I. Ogievetsky}",
    eprint = "hep-th/9812242",
    archivePrefix = "arXiv",
    reportNumber = "FIAN-TD-27-98",
    month = "12",
    year = "1998"
}

@article{Joung:2013nma,
    author = "Joung, Euihun and Taronna, Massimo",
    title = "{Cubic-interaction-induced deformations of higher-spin symmetries}",
    eprint = "1311.0242",
    archivePrefix = "arXiv",
    primaryClass = "hep-th",
    reportNumber = "AEI-2013-258",
    doi = "10.1007/JHEP03(2014)103",
    journal = "JHEP",
    volume = "03",
    pages = "103",
    year = "2014"
}

@article{Vasiliev:1992ix,
    author = "Vasiliev, Mikhail A.",
    title = "{Equations of motion for d = 3 massless fields interacting through Chern-Simons higher spin gauge fields}",
    reportNumber = "GOTEBORG-92-12",
    doi = "10.1142/S0217732392003116",
    journal = "Mod. Phys. Lett. A",
    volume = "7",
    pages = "3689--3702",
    year = "1992"
}

@article{Bengtsson:1986kh,
    author = "Bengtsson, Anders K. H. and Bengtsson, Ingemar and Linden, Noah",
    title = "{Interacting Higher Spin Gauge Fields on the Light Front}",
    reportNumber = "QMC-86-24",
    doi = "10.1088/0264-9381/4/5/028",
    journal = "Class. Quant. Grav.",
    volume = "4",
    pages = "1333",
    year = "1987"
}

@article{Metsaev:1991mt,
    author = "Metsaev, R. R.",
    title = "{Poincare invariant dynamics of massless higher spins: Fourth order analysis on mass shell}",
    doi = "10.1142/S0217732391000348",
    journal = "Mod. Phys. Lett. A",
    volume = "6",
    pages = "359--367",
    year = "1991"
}

@article{Benincasa:2007xk,
    author = "Benincasa, Paolo and Cachazo, Freddy",
    title = "{Consistency Conditions on the S-Matrix of Massless Particles}",
    eprint = "0705.4305",
    archivePrefix = "arXiv",
    primaryClass = "hep-th",
    reportNumber = "UWO-TH-07-09",
    month = "5",
    year = "2007"
}

@article{Benincasa:2011pg,
    author = "Benincasa, Paolo and Conde, Eduardo",
    title = "{Exploring the S-Matrix of Massless Particles}",
    eprint = "1108.3078",
    archivePrefix = "arXiv",
    primaryClass = "hep-th",
    doi = "10.1103/PhysRevD.86.025007",
    journal = "Phys. Rev. D",
    volume = "86",
    pages = "025007",
    year = "2012"
}

@article{Conde:2016vxs,
    author = "Conde, Eduardo and Marzolla, Andrea",
    title = "{Lorentz Constraints on Massive Three-Point Amplitudes}",
    eprint = "1601.08113",
    archivePrefix = "arXiv",
    primaryClass = "hep-th",
    doi = "10.1007/JHEP09(2016)041",
    journal = "JHEP",
    volume = "09",
    pages = "041",
    year = "2016"
}

@article{Brown:1986nw,
    author = "Brown, J. David and Henneaux, M.",
    title = "{Central Charges in the Canonical Realization of Asymptotic Symmetries: An Example from Three-Dimensional Gravity}",
    doi = "10.1007/BF01211590",
    journal = "Commun. Math. Phys.",
    volume = "104",
    pages = "207--226",
    year = "1986"
}

@article{Campoleoni:2010zq,
    author = "Campoleoni, Andrea and Fredenhagen, Stefan and Pfenninger, Stefan and Theisen, Stefan",
    title = "{Asymptotic symmetries of three-dimensional gravity coupled to higher-spin fields}",
    eprint = "1008.4744",
    archivePrefix = "arXiv",
    primaryClass = "hep-th",
    reportNumber = "AEI-2010-140",
    doi = "10.1007/JHEP11(2010)007",
    journal = "JHEP",
    volume = "11",
    pages = "007",
    year = "2010"
}

@article{Campoleoni:2011hg,
    author = "Campoleoni, Andrea and Fredenhagen, Stefan and Pfenninger, Stefan",
    title = "{Asymptotic W-symmetries in three-dimensional higher-spin gauge theories}",
    eprint = "1107.0290",
    archivePrefix = "arXiv",
    primaryClass = "hep-th",
    reportNumber = "AEI-2011-041",
    doi = "10.1007/JHEP09(2011)113",
    journal = "JHEP",
    volume = "09",
    pages = "113",
    year = "2011"
}

@article{Henneaux:2010xg,
    author = "Henneaux, Marc and Rey, Soo-Jong",
    title = "{Nonlinear $W_{infinity}$ as Asymptotic Symmetry of Three-Dimensional Higher Spin Anti-de Sitter Gravity}",
    eprint = "1008.4579",
    archivePrefix = "arXiv",
    primaryClass = "hep-th",
    doi = "10.1007/JHEP12(2010)007",
    journal = "JHEP",
    volume = "12",
    pages = "007",
    year = "2010"
}

@book{Fredenhagen:2024kqn,
    editor = "Fredenhagen, Stefan",
    title = "{Introductory Lectures on Higher-Spin Theories}",
    doi = "10.1007/978-3-031-59656-8",
    isbn = "978-3-031-59655-1, 978-3-031-59656-8",
    volume = "1028 of {\it Lecture Notes in
Physics}, Springer (2024)"
}

@article{Campoleoni:2012hp,
    author = "Campoleoni, Andrea and Fredenhagen, Stefan and Pfenninger, Stefan and Theisen, Stefan",
    title = "{Towards metric-like higher-spin gauge theories in three dimensions}",
    eprint = "1208.1851",
    archivePrefix = "arXiv",
    primaryClass = "hep-th",
    reportNumber = "AEI-2012-081",
    doi = "10.1088/1751-8113/46/21/214017",
    journal = "J. Phys. A",
    volume = "46",
    pages = "214017",
    year = "2013"
}

@article{Gwak:2015jdo,
    author = "Gwak, Seungho and Joung, Euihun and Mkrtchyan, Karapet and Rey, Soo-Jong",
    title = "{Rainbow vacua of colored higher-spin (A)dS$_{3}$ gravity}",
    eprint = "1511.05975",
    archivePrefix = "arXiv",
    primaryClass = "hep-th",
    doi = "10.1007/JHEP05(2016)150",
    journal = "JHEP",
    volume = "05",
    pages = "150",
    year = "2016"
}

@article{Alkalaev:2021zda,
    author = "Alkalaev, Konstantin and Yan, Alexander",
    title = "{AdS$_{3}$/AdS$_{2}$ degression of massless particles}",
    eprint = "2105.05722",
    archivePrefix = "arXiv",
    primaryClass = "hep-th",
    doi = "10.1007/JHEP09(2021)198",
    journal = "JHEP",
    volume = "09",
    pages = "198",
    year = "2021"
}

@article{Bekaert:2022poo,
    author = "Bekaert, Xavier and Boulanger, Nicolas and Campoleoni, Andrea and Chiodaroli, Marco and Francia, Dario and Grigoriev, Maxim and Sezgin, Ergin and Skvortsov, Evgeny",
    title = "{Snowmass White Paper: Higher Spin Gravity and Higher Spin Symmetry}",
    eprint = "2205.01567",
    archivePrefix = "arXiv",
    primaryClass = "hep-th",
    month = "5",
    year = "2022"
}

@article{Afshar:2013vka,
    author = "Afshar, Hamid and Bagchi, Arjun and Fareghbal, Reza and Grumiller, Daniel and Rosseel, Jan",
    title = "{Spin-3 Gravity in Three-Dimensional Flat Space}",
    eprint = "1307.4768",
    archivePrefix = "arXiv",
    primaryClass = "hep-th",
    reportNumber = "TUW-13-09",
    doi = "10.1103/PhysRevLett.111.121603",
    journal = "Phys. Rev. Lett.",
    volume = "111",
    number = "12",
    pages = "121603",
    year = "2013"
}

@article{Vasiliev:1989re,
    author = "Vasiliev, Mikhail A.",
    title = "{Higher Spin Algebras and Quantization on the Sphere and Hyperboloid}",
    reportNumber = "LEBEDEV-89-214",
    doi = "10.1142/S0217751X91000605",
    journal = "Int. J. Mod. Phys. A",
    volume = "6",
    pages = "1115--1135",
    year = "1991"
}

@article{Grigoriev:2020lzu,
    author = "Grigoriev, Maxim and Mkrtchyan, Karapet and Skvortsov, Evgeny",
    title = "{Matter-free higher spin gravities in 3D: Partially-massless fields and general structure}",
    eprint = "2005.05931",
    archivePrefix = "arXiv",
    primaryClass = "hep-th",
    doi = "10.1103/PhysRevD.102.066003",
    journal = "Phys. Rev. D",
    volume = "102",
    number = "6",
    pages = "066003",
    year = "2020"
}

@article{Boulanger:2023tvt,
    author = "Boulanger, Nicolas and Campoleoni, Andrea and Lekeu, Victor and Skvortsov, Evgeny",
    title = "{Strange higher-spin topological systems in 3D}",
    eprint = "2312.03382",
    archivePrefix = "arXiv",
    primaryClass = "hep-th",
    doi = "10.1007/JHEP05(2024)109",
    journal = "JHEP",
    volume = "05",
    pages = "109",
    year = "2024"
}

@misc{KingPomfretCode2026,
  author       = {King, Freddie and Pomfret, Taylor},
  title        = {{Accompanying code for `Metric-like Cubic Vertices for Massless Bosonic Higher-Spin Fields in $\mathrm{AdS}_3$'}},
  year         = {2026},
  howpublished = {\url{https://github.com/fk260/ads3_higher_spin_cubic_vertices_gauge_variation_and_DDIs}},
  note         = {Version 1.0.0}
}

@article{Beccaria:2016tqy,
    author = "Beccaria, M. and Tseytlin, A. A.",
    title = "{Iterating free-field AdS/CFT: higher spin partition function relations}",
    eprint = "1602.00948",
    archivePrefix = "arXiv",
    primaryClass = "hep-th",
    reportNumber = "IMPERIAL-TP-AT-2016-01",
    doi = "10.1088/1751-8113/49/29/295401",
    journal = "J. Phys. A",
    volume = "49",
    number = "29",
    pages = "295401",
    year = "2016"
}

@article{Beccaria:2015vaa,
    author = "Beccaria, Matteo and Tseytlin, A. A.",
    title = "{On higher spin partition functions}",
    eprint = "1503.08143",
    archivePrefix = "arXiv",
    primaryClass = "hep-th",
    reportNumber = "IMPERIAL-TP-AT-2015-01",
    doi = "10.1088/1751-8113/48/27/275401",
    journal = "J. Phys. A",
    volume = "48",
    number = "27",
    pages = "275401",
    year = "2015"
}

@article{Giombi:2014yra,
    author = "Giombi, Simone and Klebanov, Igor R. and Tseytlin, Arkady A.",
    title = "{Partition Functions and Casimir Energies in Higher Spin AdS$_{d+1}$/CFT$_d$}",
    eprint = "1402.5396",
    archivePrefix = "arXiv",
    primaryClass = "hep-th",
    reportNumber = "PUTP-2460, IMPERIAL-TP-AT-2014-01",
    doi = "10.1103/PhysRevD.90.024048",
    journal = "Phys. Rev. D",
    volume = "90",
    number = "2",
    pages = "024048",
    year = "2014"
}

@article{Gunaydin:2016amv,
    author = {G{\"u}naydin, Murat and Skvortsov, E. D. and Tran, Tung},
    title = "{Exceptional $F(4)$ higher-spin theory in AdS$_{6}$ at one-loop and other tests of duality}",
    eprint = "1608.07582",
    archivePrefix = "arXiv",
    primaryClass = "hep-th",
    reportNumber = "LMU-ASC-41-16",
    doi = "10.1007/JHEP11(2016)168",
    journal = "JHEP",
    volume = "11",
    pages = "168",
    year = "2016"
}

@article{Didenko:2014dwa,
    author = "Didenko, V. E. and Skvortsov, E. D.",
    title = "{Elements of Vasiliev Theory}",
    eprint = "1401.2975",
    archivePrefix = "arXiv",
    primaryClass = "hep-th",
    doi = "10.1007/978-3-031-59656-8_3",
    journal = "Lect. Notes Phys.",
    volume = "1028",
    pages = "269--456",
    year = "2024"
}

@article{Vasiliev:1995dn,
    author = "Vasiliev, Mikhail A.",
    editor = "Berezin, V. A. and Rubakov, V. A. and Semikoz, D. V.",
    title = "{Higher spin gauge theories in four-dimensions, three-dimensions, and two-dimensions}",
    eprint = "hep-th/9611024",
    archivePrefix = "arXiv",
    reportNumber = "FIAN-TD-24-96",
    doi = "10.1142/S0218271896000473",
    journal = "Int. J. Mod. Phys. D",
    volume = "5",
    pages = "763--797",
    year = "1996"
}

@article{Vasiliev:2003ev,
    author = "Vasiliev, M. A.",
    title = "{Nonlinear equations for symmetric massless higher spin fields in (A)dS(d)}",
    eprint = "hep-th/0304049",
    archivePrefix = "arXiv",
    reportNumber = "FIAN-TD-07-03",
    doi = "10.1016/S0370-2693(03)00872-4",
    journal = "Phys. Lett. B",
    volume = "567",
    pages = "139--151",
    year = "2003"
}

@article{Bae:2016rgm,
    author = "Bae, Jin-Beom and Joung, Euihun and Lal, Shailesh",
    title = "{One-loop test of free SU(N ) adjoint model holography}",
    eprint = "1603.05387",
    archivePrefix = "arXiv",
    primaryClass = "hep-th",
    doi = "10.1007/JHEP04(2016)061",
    journal = "JHEP",
    volume = "04",
    pages = "061",
    year = "2016"
}

@article{Bae:2016hfy,
    author = "Bae, Jin-Beom and Joung, Euihun and Lal, Shailesh",
    title = "{On the Holography of Free Yang-Mills}",
    eprint = "1607.07651",
    archivePrefix = "arXiv",
    primaryClass = "hep-th",
    doi = "10.1007/JHEP10(2016)074",
    journal = "JHEP",
    volume = "10",
    pages = "074",
    year = "2016"
}

@article{Bae:2017spv,
    author = "Bae, Jin-Beom and Joung, Euihun and Lal, Shailesh",
    title = "{One-loop free energy of tensionless type IIB string in AdS$_5\times$S$^5$}",
    eprint = "1701.01507",
    archivePrefix = "arXiv",
    primaryClass = "hep-th",
    doi = "10.1007/JHEP06(2017)155",
    journal = "JHEP",
    volume = "06",
    pages = "155",
    year = "2017"
}

@article{Bae:2017fcs,
    author = "Bae, Jin-Beom and Joung, Euihun and Lal, Shailesh",
    title = "{Exploring Free Matrix CFT Holographies at One-Loop}",
    eprint = "1708.04644",
    archivePrefix = "arXiv",
    primaryClass = "hep-th",
    doi = "10.3390/universe3040077",
    journal = "Universe",
    volume = "3",
    number = "4",
    pages = "77",
    year = "2017"
}

@article{Basile:2018zoy,
    author = "Basile, Thomas and Joung, Euihun and Lal, Shailesh and Li, Wenliang",
    title = "{Character Integral Representation of Zeta function in AdS$_{d+1}$: I. Derivation of the general formula}",
    eprint = "1805.05646",
    archivePrefix = "arXiv",
    primaryClass = "hep-th",
    doi = "10.1007/JHEP10(2018)091",
    journal = "JHEP",
    volume = "10",
    pages = "091",
    year = "2018"
}

@article{Konstein:2000bi,
    author = "Konstein, S. E. and Vasiliev, M. A. and Zaikin, V. N.",
    title = "{Conformal higher spin currents in any dimension and AdS / CFT correspondence}",
    eprint = "hep-th/0010239",
    archivePrefix = "arXiv",
    reportNumber = "FIAN-TD-16-00",
    doi = "10.1088/1126-6708/2000/12/018",
    journal = "JHEP",
    volume = "12",
    pages = "018",
    year = "2000"
}

@article{Didenko:2017lsn,
    author = "Didenko, V. E. and Vasiliev, M. A.",
    title = "{Test of the local form of higher-spin equations via AdS / CFT}",
    eprint = "1705.03440",
    archivePrefix = "arXiv",
    primaryClass = "hep-th",
    reportNumber = "FIAN-TD-10-17",
    doi = "10.1016/j.physletb.2017.09.091",
    journal = "Phys. Lett. B",
    volume = "775",
    pages = "352--360",
    year = "2017"
}

@article{Giombi:2013yva,
    author = "Giombi, Simone and Klebanov, Igor R. and Pufu, Silviu S. and Safdi, Benjamin R. and Tarnopolsky, Grigory",
    title = "{AdS Description of Induced Higher-Spin Gauge Theory}",
    eprint = "1306.5242",
    archivePrefix = "arXiv",
    primaryClass = "hep-th",
    reportNumber = "MIT-CTP-4471",
    doi = "10.1007/JHEP10(2013)016",
    journal = "JHEP",
    volume = "10",
    pages = "016",
    year = "2013"
}

@article{Giombi:2013fka,
    author = "Giombi, Simone and Klebanov, Igor R.",
    title = "{One Loop Tests of Higher Spin AdS/CFT}",
    eprint = "1308.2337",
    archivePrefix = "arXiv",
    primaryClass = "hep-th",
    reportNumber = "PUTP-2451",
    doi = "10.1007/JHEP12(2013)068",
    journal = "JHEP",
    volume = "12",
    pages = "068",
    year = "2013"
}

@article{Giombi:2014iua,
    author = "Giombi, Simone and Klebanov, Igor R. and Safdi, Benjamin R.",
    title = "{Higher Spin AdS$_{d+1}$/CFT$_d$ at One Loop}",
    eprint = "1401.0825",
    archivePrefix = "arXiv",
    primaryClass = "hep-th",
    reportNumber = "PUPT-2458",
    doi = "10.1103/PhysRevD.89.084004",
    journal = "Phys. Rev. D",
    volume = "89",
    number = "8",
    pages = "084004",
    year = "2014"
}

@inproceedings{Giombi:2016ejx,
    author = "Giombi, Simone",
    title = "{Higher Spin {\textemdash} CFT Duality}",
    booktitle = "{Theoretical Advanced Study Institute in Elementary Particle Physics}: {New Frontiers in Fields and Strings}",
    eprint = "1607.02967",
    archivePrefix = "arXiv",
    primaryClass = "hep-th",
    doi = "10.1142/9789813149441_0003",
    pages = "137--214",
    year = "2017"
}

@article{Giombi:2016pvg,
    author = "Giombi, Simone and Klebanov, Igor R. and Tan, Zhong Ming",
    title = "{The ABC of Higher-Spin AdS/CFT}",
    eprint = "1608.07611",
    archivePrefix = "arXiv",
    primaryClass = "hep-th",
    reportNumber = "PUPT-2505",
    doi = "10.3390/universe4010018",
    journal = "Universe",
    volume = "4",
    number = "1",
    pages = "18",
    year = "2018"
}

@article{Giombi:2016zwa,
    author = "Giombi, S. and Gurucharan, V. and Kirilin, V. and Prakash, S. and Skvortsov, E.",
    title = "{On the Higher-Spin Spectrum in Large N Chern-Simons Vector Models}",
    eprint = "1610.08472",
    archivePrefix = "arXiv",
    primaryClass = "hep-th",
    reportNumber = "PUPT-2512, LMU-ASC-52-16",
    doi = "10.1007/JHEP01(2017)058",
    journal = "JHEP",
    volume = "01",
    pages = "058",
    year = "2017"
}

@article{Manvelyan:2004ii,
    author = "Manvelyan, Ruben and Ruhl, Werner",
    title = "{The Masses of gauge fields in higher spin field theory on the bulk of AdS(4)}",
    eprint = "hep-th/0412252",
    archivePrefix = "arXiv",
    doi = "10.1016/j.physletb.2005.03.061",
    journal = "Phys. Lett. B",
    volume = "613",
    pages = "197--207",
    year = "2005"
}

@article{Manvelyan:2008ks,
    author = "Manvelyan, Ruben and Mkrtchyan, Karapet and Ruhl, Werner",
    title = "{Ultraviolet behaviour of higher spin gauge field propagators and one loop mass renormalization}",
    eprint = "0804.1211",
    archivePrefix = "arXiv",
    primaryClass = "hep-th",
    doi = "10.1016/j.nuclphysb.2008.06.008",
    journal = "Nucl. Phys. B",
    volume = "803",
    pages = "405--427",
    year = "2008"
}

@article{Maldacena:2011jn,
    author = "Maldacena, Juan and Zhiboedov, Alexander",
    title = "{Constraining Conformal Field Theories with A Higher Spin Symmetry}",
    eprint = "1112.1016",
    archivePrefix = "arXiv",
    primaryClass = "hep-th",
    doi = "10.1088/1751-8113/46/21/214011",
    journal = "J. Phys. A",
    volume = "46",
    pages = "214011",
    year = "2013"
}

@article{Sleight:2016dba,
    author = "Sleight, Charlotte and Taronna, Massimo",
    title = "{Higher Spin Interactions from Conformal Field Theory: The Complete Cubic Couplings}",
    eprint = "1603.00022",
    archivePrefix = "arXiv",
    primaryClass = "hep-th",
    reportNumber = "MPP-2016-25",
    doi = "10.1103/PhysRevLett.116.181602",
    journal = "Phys. Rev. Lett.",
    volume = "116",
    number = "18",
    pages = "181602",
    year = "2016"
}

@article{Bekaert:2014cea,
    author = "Bekaert, Xavier and Erdmenger, Johanna and Ponomarev, Dmitry and Sleight, Charlotte",
    title = "{Towards holographic higher-spin interactions: Four-point functions and higher-spin exchange}",
    eprint = "1412.0016",
    archivePrefix = "arXiv",
    primaryClass = "hep-th",
    doi = "10.1007/JHEP03(2015)170",
    journal = "JHEP",
    volume = "03",
    pages = "170",
    year = "2015"
}

@article{Fredenhagen:2019hvb,
    author = {Fredenhagen, Stefan and Kr{\"u}ger, Olaf and Mkrtchyan, Karapet},
    title = "{Vertex-Constraints in 3D Higher Spin Theories}",
    eprint = "1905.00093",
    archivePrefix = "arXiv",
    primaryClass = "hep-th",
    doi = "10.1103/PhysRevLett.123.131601",
    journal = "Phys. Rev. Lett.",
    volume = "123",
    number = "13",
    pages = "131601",
    year = "2019"
}

@article{deMelloKoch:2014vnt,
    author = "de Mello Koch, Robert and Jevicki, Antal and Rodrigues, Jo{\~a}o P. and Yoon, Junggi",
    title = "{Canonical Formulation of $O(N)$ Vector/Higher Spin Correspondence}",
    eprint = "1408.4800",
    archivePrefix = "arXiv",
    primaryClass = "hep-th",
    reportNumber = "BROWN-HET-1659, WITS-CTP-146",
    doi = "10.1088/1751-8113/48/10/105403",
    journal = "J. Phys. A",
    volume = "48",
    number = "10",
    pages = "105403",
    year = "2015"
}

@article{Aharony:2020omh,
    author = "Aharony, Ofer and Chester, Shai M. and Urbach, Erez Y.",
    title = "{A Derivation of AdS/CFT for Vector Models}",
    eprint = "2011.06328",
    archivePrefix = "arXiv",
    primaryClass = "hep-th",
    doi = "10.1007/JHEP03(2021)208",
    journal = "JHEP",
    volume = "03",
    pages = "208",
    year = "2021"
}

@article{Costa:2014kfa,
    author = "Costa, Miguel S. and Gon{\c{c}}alves, Vasco and Penedones, Jo{\~a}o",
    title = "{Spinning AdS Propagators}",
    eprint = "1404.5625",
    archivePrefix = "arXiv",
    primaryClass = "hep-th",
    doi = "10.1007/JHEP09(2014)064",
    journal = "JHEP",
    volume = "09",
    pages = "064",
    year = "2014"
}

@article{Neiman:2022enh,
    author = "Neiman, Yasha",
    title = "{New Diagrammatic Framework for Higher-Spin Gravity}",
    eprint = "2209.02185",
    archivePrefix = "arXiv",
    primaryClass = "hep-th",
    doi = "10.1103/PhysRevLett.130.171601",
    journal = "Phys. Rev. Lett.",
    volume = "130",
    number = "17",
    pages = "171601",
    year = "2023"
}

@article{Fredenhagen:2018guf,
    author = {Fredenhagen, Stefan and Kr{\"u}ger, Olaf and Mkrtchyan, Karapet},
    title = "{Constraints for Three-Dimensional Higher-Spin Interactions and Conformal Correlators}",
    eprint = "1812.10462",
    archivePrefix = "arXiv",
    primaryClass = "hep-th",
    doi = "10.1103/PhysRevD.100.066019",
    journal = "Phys. Rev. D",
    volume = "100",
    number = "6",
    pages = "066019",
    year = "2019"
}

@article{Vasiliev:1986bq,
    author = "Vasiliev, Mikhail A. and Fradkin, E. S.",
    title = "{Gravitational Interaction of Massless High Spin (S {\ensuremath{>}} 2) Fields}",
    journal = "JETP Lett.",
    volume = "44",
    pages = "622--627",
    year = "1986"
}

@article{Fradkin:1987ks,
    author = "Fradkin, E. S. and Vasiliev, Mikhail A.",
    title = "{On the Gravitational Interaction of Massless Higher Spin Fields}",
    doi = "10.1016/0370-2693(87)91275-5",
    journal = "Phys. Lett. B",
    volume = "189",
    pages = "89--95",
    year = "1987"
}

@article{Fradkin:1986qy,
    author = "Fradkin, E. S. and Vasiliev, Mikhail A.",
    title = "{Cubic Interaction in Extended Theories of Massless Higher Spin Fields}",
    reportNumber = "LEBEDEV-86-309",
    doi = "10.1016/0550-3213(87)90469-X",
    journal = "Nucl. Phys. B",
    volume = "291",
    pages = "141--171",
    year = "1987"
}

@article{Pekar:2023nev,
    author = "Pekar, Simon",
    title = "{Introduction to higher-spin theories}",
    doi = "10.22323/1.435.0004",
    journal = "PoS",
    volume = "Modave2022",
    pages = "004",
    year = "2023"
}

@inproceedings{Porrati:2012rd,
    author = "Porrati, M.",
    title = "{Old and New No Go Theorems on Interacting Massless Particles in Flat Space}",
    booktitle = "{17th International Seminar on High Energy Physics}",
    eprint = "1209.4876",
    archivePrefix = "arXiv",
    primaryClass = "hep-th",
    month = "9",
    year = "2012"
}

@article{Metsaev:2005ar,
    author = "Metsaev, R. R.",
    title = "{Cubic interaction vertices of massive and massless higher spin fields}",
    eprint = "hep-th/0512342",
    archivePrefix = "arXiv",
    reportNumber = "FIAN-TD-19-05",
    doi = "10.1016/j.nuclphysb.2006.10.002",
    journal = "Nucl. Phys. B",
    volume = "759",
    pages = "147--201",
    year = "2006"
}

@article{Metsaev:2007rn,
    author = "Metsaev, R. R.",
    title = "{Cubic interaction vertices for fermionic and bosonic arbitrary spin fields}",
    eprint = "0712.3526",
    archivePrefix = "arXiv",
    primaryClass = "hep-th",
    reportNumber = "FIAN-TD-2007-25",
    doi = "10.1016/j.nuclphysb.2012.01.022",
    journal = "Nucl. Phys. B",
    volume = "859",
    pages = "13--69",
    year = "2012"
}

@article{Bengtsson:1983pd,
    author = "Bengtsson, Anders K. H. and Bengtsson, Ingemar and Brink, Lars",
    title = "{Cubic Interaction Terms for Arbitrary Spin}",
    reportNumber = "GOTEBORG-83-10",
    doi = "10.1016/0550-3213(83)90140-2",
    journal = "Nucl. Phys. B",
    volume = "227",
    pages = "31--40",
    year = "1983"
}

@article{Khabarov:2020bgr,
    author = "Khabarov, M. V. and Zinoviev, Yu. M.",
    title = "{Massless higher spin cubic vertices in flat four dimensional space}",
    eprint = "2005.09851",
    archivePrefix = "arXiv",
    primaryClass = "hep-th",
    doi = "10.1007/JHEP08(2020)112",
    journal = "JHEP",
    volume = "08",
    pages = "112",
    year = "2020"
}

@article{Fierz:1939zz,
    author = "Fierz, M.",
    title = "{Force-free particles with any spin}",
    journal = "Helv. Phys. Acta",
    volume = "12",
    pages = "3--37",
    year = "1939"
}

@article{Giombi:2011ya,
    author = "Giombi, Simone and Yin, Xi",
    title = "{On Higher Spin Gauge Theory and the Critical O(N) Model}",
    eprint = "1105.4011",
    archivePrefix = "arXiv",
    primaryClass = "hep-th",
    doi = "10.1103/PhysRevD.85.086005",
    journal = "Phys. Rev. D",
    volume = "85",
    pages = "086005",
    year = "2012"
}

@article{Vasiliev:2011knf,
    author = "Vasiliev, M. A.",
    title = "{Cubic Vertices for Symmetric Higher-Spin Gauge Fields in $(A)dS_d$}",
    eprint = "1108.5921",
    archivePrefix = "arXiv",
    primaryClass = "hep-th",
    reportNumber = "FIAN-TD-11-17",
    doi = "10.1016/j.nuclphysb.2012.04.012",
    journal = "Nucl. Phys. B",
    volume = "862",
    pages = "341--408",
    year = "2012"
}

@article{Metsaev:2018xip,
    author = "Metsaev, R. R.",
    title = "{Light-cone gauge cubic interaction vertices for massless fields in AdS(4)}",
    eprint = "1807.07542",
    archivePrefix = "arXiv",
    primaryClass = "hep-th",
    reportNumber = "FIAN-TD-2018-15",
    doi = "10.1016/j.nuclphysb.2018.09.021",
    journal = "Nucl. Phys. B",
    volume = "936",
    pages = "320--351",
    year = "2018"
}

@article{Boulanger:2012dx,
    author = "Boulanger, Nicolas and Ponomarev, Dmitry and Skvortsov, E. D.",
    title = "{Non-abelian cubic vertices for higher-spin fields in anti-de Sitter space}",
    eprint = "1211.6979",
    archivePrefix = "arXiv",
    primaryClass = "hep-th",
    doi = "10.1007/JHEP05(2013)008",
    journal = "JHEP",
    volume = "05",
    pages = "008",
    year = "2013"
}
\end{document}